\documentclass[10pt,journal]{IEEEtran}
\ifCLASSOPTIONcompsoc
  \usepackage[nocompress]{cite}
\else
  \usepackage{cite}
\fi
\ifCLASSINFOpdf
\else
\fi
\usepackage{stfloats}
\usepackage{graphicx}
\usepackage{booktabs}
\usepackage{float}
\usepackage[justification=centering]{caption}
\usepackage{amssymb,amsfonts}
\usepackage{amsmath}
\usepackage{listings}
\usepackage{url}
\usepackage{algorithm,algorithmicx}
\usepackage{algpseudocode}

\begin{document}
\title{Fast Rigid 3D Registration Solution: A Simple Method Free of SVD and Eigen-Decomposition}

\author{Jin~Wu,~\IEEEmembership{Member,~IEEE,}
            Ming~Liu,~\IEEEmembership{Member,~IEEE,}
            Zebo~Zhou
            and~Rui~Li,~\IEEEmembership{Member,~IEEE}
\thanks{This research was supported by General Research Fund of Research Grants Council Hong Kong,11210017, in part by Early Career Scheme Project of Research Grants Council Hong Kong, 21202816 and in part by National Natural Science Foundation of China with the grant of No. 41604025. Corresponding author: Ming Liu.}
\thanks{J. Wu is with Department of Electronic \& Computer Engineering, Hong Kong University of Science and Technology, Hong Kong, China and is also with School of Automation, University of Electronic Science and Technology of China, Chengdu, 611731, China. e-mail: jin\_wu\_uestc@hotmail.com.}
\thanks {M. Liu is with Department of Electronic \& Computer Engineering, Hong Kong University of Science and Technology, Hong Kong, China. e-mail: eelium@ust.hk; liu.ming.prc@gmail.com.}
\thanks {Z. Zhou is with School of Aeronautics and Astronautics, University of Electronic Science and Technology of China, Chengdu, 611731, China. e-mail: klinsmann.zhou@gmail.com.}
\thanks {R. Li is with School of Automation, University of Electronic Science and Technology of China, Chengdu, 611731, China. e-mail: hitlirui@gmail.com.}
\thanks{The source codes of the proposed algorithm and comparisons can be downloaded on https://github.com/zarathustr/FA3R.}
%\IEEEcompsocthanksitem J. Kittler is with Centre for Vision, Speech and Signal Processing, Faculty of Engineering and Physical Sciences, University of Surrey, Guildford, UK. e-mail: 
}

\IEEEtitleabstractindextext{%
\begin{abstract}
A novel solution is obtained to solve the rigid 3D registration problem, motivated by previous eigen-decomposition approaches. Different from existing solvers, the proposed algorithm does not require sophisticated matrix operations e.g. singular value decomposition or eigenvalue decomposition. Instead, the optimal eigenvector of the point cross-covariance matrix can be computed within several iterations. It is also proven that the optimal rotation matrix can be directly computed for cases without need of quaternion. The simple framework provides very easy approach of integer-implementation on embedded platforms. Simulations on noise-corrupted point clouds have verified the robustness and computation speed of the proposed method. The final results indicate that the proposed algorithm is accurate, robust and owns over $60\% \sim 80\%$ less computation time than representatives. It has also been applied to real-world applications for faster relative robotic navigation.
\end{abstract}

}

\maketitle

\IEEEdisplaynontitleabstractindextext

% \fi
\IEEEpeerreviewmaketitle

\section{Introduction}\label{sec:introduction}

\IEEEPARstart{3}{D} registration from point clouds is a crucial factor in assembly automation and robotics \cite{Aghili2016,Liu2014}. Using scanner or cameras, the point vector measurement from various directions can be captured to reconstruct the object's shape by cloud matching \cite{Liu2016,Ye2015}. The rigid 3D registration achieves this by estimating the rotation and translation between two corresponding point sets.\\
\indent The 3D registration has been studied for quite a long time. In early 1980s, the scientists tried to extract motions from the medical images eager to have better multidimensional understanding on the surgery \cite{faber1988orientation,venot1984new}. Some pioneers employ tensor algebra to estimate the transformation parameters \cite{Cyganski1986}. The batch-processing of 3D registration is regarded to be first successfully solved by Arun et al. in 1987 \cite{Arun1987} with the singular value decomposition (SVD),  although there have been some similar early ideas occured in very early solutions to Wahba's problem since 1960s \cite{Wahba1965, Davenport1968}. Arun's method is not robust enough and it has been improved by Umeyama in the later 4 years \cite{Umeyama1991}. The accurate estimation by means of SVD allows for very fast computation of orientation and translation which are highly needed by the robotic devices. The idea of the closest iterative points (ICP, \cite{Besl1992, Zhang1994}) was then proposed to considering the registration of two point sets with different dimensions i.e. numbers of points. In the conventional ICP literature \cite{Besl1992}, the eigenvalue decomposition (or eigen-decomposition i.e. EIG) method is introduced where the rotation matrix is parameterized via the unit quaternion, which has also been studied early by Horn in 1987 \cite{Horn1987}. With the creation of ICP, the 3D registration has welcomed its huge boost in industrial applications. According to the main lacks of ICP i.e. the convergence and local optimum, there have been over 100 variants of it trying to improve the practical performance \cite{Jiaolong2016}. A very recent research has also indicated possible further essence of the ICP \cite{Kwok2018}.

\setcounter{equation}{9}
\begin{figure*}[hb]
\begin{equation}\label{G}
{\mathbf{G}} = \left( {\begin{array}{*{20}{c}}
  {{X_{xx}} + {X_{yy}} + {X_{zz}}}&{{X_{yz}} - {X_{zy}}}&{{X_{zx}} - {X_{xz}}}&{{X_{xy}} - {X_{yx}}} \\ 
  {{X_{yz}} - {X_{zy}}}&{{X_{xx}} - {X_{yy}} - {X_{zz}}}&{{X_{xy}} + {X_{yx}}}&{{X_{xz}} + {X_{zx}}} \\ 
  {{X_{zx}} - {X_{xz}}}&{{X_{xy}} + {X_{yx}}}&{ - {X_{xx}} + {X_{yy}} - {X_{zz}}}&{{X_{yz}} + {X_{zy}}} \\ 
  {{X_{xy}} - {X_{yx}}}&{{X_{xz}} + {X_{zx}}}&{{X_{yz}} + {X_{zy}}}&{ - {X_{x1}} - {X_{yy}} + {X_{zz}}} 
\end{array}} \right)
\end{equation}
\end{figure*}

\indent Apart from the SVD and EIG, the 3D registration is also accomplished by other methods e.g. dual-quaternion algorithm by Wakler et al. \cite{Walker1991}. Besides, as 3D registration is described as a least-square problem, it has also been solved by gradient-descent algorithm (GDA) and Levenberg-Marquardt algorithm (LMA). The robust 3D registration has also been discussed deeply by \cite{fitzgibbon2003robust,jian2011robust} as in real engineering scenarios there are too many local optimum setting obstacles for global searching optimality. During the past 20 years, the kernel problem of ICP is usually dealt with by SVD or EIG since they are evaluated to be accurate, fast, intuitive and easy-to-implement \cite{eggert1997estimating}. For nowadays open-source libraries, most of the tasks are completed by the two tools as well. For instance, in the ETHZ-ASL 'libpointmatcher' library and the latest public codes of Go-ICP \cite{Jiaolong2016}, the rigid transformation is computed by SVD. There has been no faster solver for over 20 years based on this classical problem. The SVD and EIG are usually implemented on the computer with numerical algorithms, which require high loads of floating-point processings. In nowadays, there are many parallel computing tools for accelerated performance of point-cloud registration e.g. the field programmable gate array (FPGA) and graphics processing unit (GPU). Such platforms can not directly process the floating-point numbers. Instead, numerical algorithms are implemented using the fixed-point integers in a sophisticated way which also adds instability to the designed system. Therefore, to boost the algorithmic computation speed on such platform, a novel method should be developed whereas the scheme must be extremely simple and intuitive. Would there be a possibility that this estimation may run even faster? The answer is positive. In this paper, motivated by previous algorithms, we derive a new algorithm in which the matrix manipulation results are rearranged properly. By analyzing the matrix power, an iterative rule is established where only simple linear vector algebra exists. The proposed method is then verified to have much less computation time than those consumed by SVD, EIG and their improved versions.\\
\indent This paper is structured as follows: Section II contains problem formulation, backgrounds, equivalence analysis and our proposed solution. In Section III the simulation results are presented to demonstrate the efficacy of the proposed algorithm. The concluding remarks are drawn in the Section IV.\\

\begin{figure*}[ht]
\begin{equation}\label{W}
{\mathbf{W}} = \left( {\begin{array}{*{20}{c}}
  {{H_{x1}} + {H_{y2}} + {H_{z3}}}&{ - {H_{y3}} + {H_{z2}}}&{ - {H_{z1}} + {H_{x3}}}&{ - {H_{x2}} + {H_{y1}}} \\ 
  { - {H_{y3}} + {H_{z2}}}&{{H_{x1}} - {H_{y2}} - {H_{z3}}}&{{H_{x2}} + {H_{y1}}}&{{H_{x3}} + {H_{z1}}} \\ 
  { - {H_{z1}} + {H_{x3}}}&{{H_{x2}} + {H_{y1}}}&{{H_{y2}} - {H_{x1}} - {H_{z3}}}&{{H_{y3}} + {H_{z2}}} \\ 
  { - {H_{x2}} + {H_{y1}}}&{{H_{x3}} + {H_{z1}}}&{{H_{y3}} + {H_{z2}}}&{{H_{z3}} - {H_{y2}} - {H_{x1}}} 
\end{array}} \right)
\end{equation}
\end{figure*}

\section{Problem Formulation and Solution}
\subsection{Basic Framework and Equivalence}

The optimal rigid 3D registration problem from vector measurements can be characterized with \cite{Arun1987}
\setcounter{equation}{0}
\begin{equation}
L({\bf{C, T}}) = \mathop {\arg \min }\limits_{{\mathbf{C}}{{\mathbf{C}}^T} = {{\mathbf{C}}^T}{\mathbf{C}} = {\mathbf{I}}} \sum\limits_{i = 1}^n {{a_i}{{\left\| {{{\mathbf{b}}_i} - {\mathbf{C}}{{\mathbf{r}}_i}} - {\bf{T}}\right\|}^2}} 
\end{equation}
where $a_i$ denotes the positive weight  of $i$-th point pair; ${\mathbf{C}} \in SO(3)$ is the direction cosine matrix (DCM) describing the 3D rotation; $\bf{T} \in {\mathbb{R}^3}$ denotes the translational vector; ${{\mathbf{b}}_i} = {({b_{x,i}},{b_{y,i}},{b_{z,i}})^T} \in \left\{ {\cal{B}} \right\}$ and ${{\bf{r}}_i} = {({r_{x,i}},{r_{y,i}},{r_{z,i}})^T} \in \left\{ {\cal{R}} \right\}$ represent the $i$-th pair of point correspondences in body frame and reference frame, respectively. $L$ is called the metric error function that owns independent variables of $\bf{C}$ and $\bf{T}$ to be estimated. The rigid 3D registration seeks the optimal DCM and translation vector to minimize the metric error. When there are numerous point pairs, the weights can hardly be pre-determined, the above optimization will become
\begin{equation}\label{opt}
L({\bf{C, T}}) = \mathop {\arg \min }\limits_{{\mathbf{C}}{{\mathbf{C}}^T} = {{\mathbf{C}}^T}{\mathbf{C}} = {\mathbf{I}}} \sum\limits_{i = 1}^n {\frac{1}{n}{{\left\| {{{\mathbf{b}}_i} - {\mathbf{C}}{{\mathbf{r}}_i} - {\mathbf{T}}} \right\|}^2}} 
\end{equation}
by equalizing the weights. Many famous solutions have been developed to solve this optimization problem. The studied optimization (\ref{opt}) is seemingly a total least-square problem where both reference and transformed observations contain noises. In a recent work \cite{Chang2015}, it, however, has been proven to be equivalent to the classical least square. The following a main representative for solving 3D registration:
\begin{enumerate}
\item The SVD method by Arun can adequately solve most problems while Umeyama proposed an improvement for better robustness when the point measurements are severely corrupted by noises \cite{Umeyama1991}. The SVD is performed on the point-set cross-covariance matrix given by \cite{Kanatani1994}
\begin{equation}
{\mathbf{D}} = \operatorname{cov} ({\cal{R}},{\cal{B}}) = \frac{1}{n}\sum\limits_{i = 1}^n {\left( {{\mathbf{r}}_i^{} - {\mathbf{\bar r}}} \right){{\left( {{\mathbf{b}}_i^{} - {\mathbf{\bar b}}} \right)}^T}} 
\end{equation}
such that
\begin{equation}
{\mathbf{D}} = {\mathbf{US}}{{\mathbf{V}}^T}
\end{equation}
The rotation matrix ${\mathbf{C}} = {\mathbf{V}}{{\mathbf{U}}^T}$ when $\det \left( {\mathbf{U}} \right)\det \left( {\mathbf{V}} \right) =  + 1$ and for the case of $\det \left( {\mathbf{U}} \right)\det \left( {\mathbf{V}} \right) =  - 1$, the optimal solution is ${\mathbf{C}} = {\mathbf{V}}diag(1,1, - 1){{\mathbf{U}}^T}$. 

\item Another common solution is by parameterizing $\bf{C}$ with quaternion ${\mathbf{q}} = {\left( {{q_0},{q_1},{q_2},{q_3}} \right)^T}$ where $\left\| {\mathbf{q}} \right\| = 1$ ensures the norm-2 of $\bf{q}$ is unitary. Then the core problem is to find out the eigenvector associated with the maximum eigenvalue of the following symmetric matrix \cite{Aghili2016}
\begin{equation}
\begin{gathered}
  {\mathbf{Pq}} = {\lambda _{\max ,{\mathbf{P}}}}{\mathbf{q}} \hfill \\
  {\mathbf{P}} = \left[ {\begin{array}{*{20}{c}}
  {tr({\mathbf{D}})}&{{{\mathbf{d}}^T}} \\ 
  {\mathbf{d}}&{{\mathbf{D}} + {{\mathbf{D}}^T} - tr({\mathbf{D}}){\mathbf{I}}} 
\end{array}} \right] \hfill \\ 
\end{gathered} 
\end{equation}
in which
\begin{equation}
{\mathbf{d}} = {\left[ {{D_{23}} - {D_{32}},{D_{31}} - {D_{13}},{D_{12}} - {D_{21}}} \right]^T}
\end{equation}
and ${\lambda _{\max ,{\mathbf{P}}}}$ denotes the maximum eigenvalue of $\bf{P}$. 
\end{enumerate}
Although many other algorithms e.g. dual-quaternion method, orthonormal-matrix method are effective, the SVD and EIG ones are regarded to have quite similar accuracy and fastest execution speed \cite{eggert1997estimating}. For human being, the EIG is seemingly more intuitive to compute than the SVD as the eigenvalues can be found out directly from algebraic roots to the quartic characteristic polynomial. Therefore, in this paper, we focus on solving the problem by means of analytical EIG.\\
\indent For EIG, there have been also other several famous equivalent algorithms:
\begin{enumerate}
\item The eigenvalue problem has also been derived by Horn in 1987 generating the following matrix form
\begin{equation}
{\mathbf{Gq}} = {\lambda _{\max ,{\mathbf{G}}}}{\mathbf{q}}
\end{equation}
where $\bf{G}$ is given by (\ref{G}) in which
\begin{equation}
\begin{gathered}
  {\mathbf{X}} = \left( {\begin{array}{*{20}{c}}
  {{X_{xx}}}&{{X_{xy}}}&{{X_{xz}}} \\ 
  {{X_{yx}}}&{{X_{yy}}}&{{X_{yz}}} \\ 
  {{X_{zx}}}&{{X_{zy}}}&{{X_{zz}}} 
\end{array}} \right) \hfill \\
  \begin{array}{*{20}{c}}
  {}&{}&{}&{} 
\end{array} = \frac{1}{n}\sum\limits_{i = 1}^n {\left( {{\mathbf{b}}_i^{} - {\mathbf{\bar b}}} \right){{\left( {{\mathbf{r}}_i^{} - {\mathbf{\bar r}}} \right)}^T} = {{\mathbf{D}}^T}}  \hfill \\ 
\end{gathered} 
\end{equation}
\item In aerospace engineering, the optimal attitude determination from vector observations is usually obtained by a famous approach i.e. the Davenport q-method \cite{Davenport1968}, where the vector pairs are normalized and the translation vector is not estimated, such that
\begin{equation}\label{Wahba}
\mathop {\arg \min }\limits_{{\mathbf{C}}{{\mathbf{C}}^T} = {{\mathbf{C}}^T}{\mathbf{C}} = {\mathbf{I}}} \sum\limits_{i = 1}^n {{a_i}{{\left\| {{{\mathbf{b}}_i} - {\mathbf{C}}{{\mathbf{r}}_i}} \right\|}^2},s.t.\left\| {{{\mathbf{b}}_i}} \right\| = } \left\| {{{\mathbf{r}}_i}} \right\| = 1
\end{equation}
This is solved by finding the largest positive eigenvalue of the Davenport matrix $\bf{K}$ given below \cite{Yang2013}
\setcounter{equation}{11}
\begin{equation}\label{Kq_q}
\begin{array}{l}
{\bf{Kq}} = {\lambda _{\max, {\bf{K}} }}{\bf{q}}\\
{\bf{K}} = \left[ {\begin{array}{*{20}{c}}
{{\bf{B}} + {{\bf{B}}^T} - tr({\bf{B}}){\bf{I}}}&{\bf{z}}\\
{{{\bf{z}}^T}}&{tr({\bf{B}})}
\end{array}} \right]\\
{\bf{B}} = \sum\limits_{i = 1}^n {{a_i}{{\bf{b}}_i}{\bf{r}}_i^T} \\
{\bf{z}} = \sum\limits_{i = 1}^n {{a_i}{{\bf{b}}_i} \times {{\bf{r}}_i}} 
\end{array}
\end{equation}
where $\lambda _{\rm{max}, {\bf{K}}}$ is the largest eigenvalue of $\bf{K}$. In existing methods e.g. QUaternion ESTimator (QUEST, \cite{Shuster1981}), Fast Optimal Attitude Matrix (FOAM, \cite{Markley1993}) and The EStimator of the Optimal Quaternion (ESOQ, \cite{Mortari1997}), researchers focus on deriving the closed form of the characteristic polynomial of $\bf{K}$. When the positive weights $a_i, i=1,2,\cdots$ are properly normalized i.e.
\begin{equation}
\sum\limits_{i = 1}^n {{a_i}}  = 1
\end{equation}
, it has also been rigorously derived that ${\lambda _{\max ,{\mathbf{K}}}} < 1$ which allows for numerical iteration solutions of eigenvalue from the start point of $1$ \cite{Shuster1981}. 

\item Recently, a Fast Linear Attitude Estimator (FLAE, \cite{Wu2018}) has been proposed by us that is proven to be faster than these representative attitude solvers. FLAE aims to compute the eigenvalue of a $4\times4$ matrix $\bf{W}$ that is closest to 1, such that
\begin{equation}\label{Wq_q}
{\bf{Wq}} = {\lambda _{\bf{W}}}{\bf{q}}
\end{equation}
where $\bf{W}$ is in (\ref{W}) and the parameters are given by
\begin{equation}
\begin{small}
\begin{gathered}
\left( {{H_{x1}},{H_{x2}},{H_{x3}}} \right) \hfill \\
  \begin{array}{*{20}{c}}
  {}&{} 
\end{array} = \left( {\begin{array}{*{20}{c}}
  {\sum\limits_{i = 1}^n {{a_i}r_{x,i}^{}b_{x,i}^{}} }&{\sum\limits_{i = 1}^n {{a_i}r_{x,i}^{}b_{y,i}^{}} }&{\sum\limits_{i = 1}^n {{a_i}r_{x,i}^{}b_{z,i}^{}} } 
\end{array}} \right) \hfill \\
 \left( {{H_{y1}},{H_{y2}},{H_{y3}}} \right) \hfill \\
  \begin{array}{*{20}{c}}
  {}&{} 
\end{array} = \left( {\begin{array}{*{20}{c}}
  {\sum\limits_{i = 1}^n {{a_i}r_{y,i}^{}b_{x,i}^{}} }&{\sum\limits_{i = 1}^n {{a_i}r_{y,i}^{}b_{y,i}^{}} }&{\sum\limits_{i = 1}^n {{a_i}r_{y,i}^{}b_{z,i}^{}} } 
\end{array}} \right) \hfill \\
\left( {{H_{z1}},{H_{z2}},{H_{z3}}} \right) \hfill \\
  \begin{array}{*{20}{c}}
  {}&{} 
\end{array} = \left( {\begin{array}{*{20}{c}}
  {\sum\limits_{i = 1}^n {{a_i}r_{z,i}^{}b_{x,i}^{}} }&{\sum\limits_{i = 1}^n {{a_i}r_{z,i}^{}b_{y,i}^{}} }&{\sum\limits_{i = 1}^n {{a_i}r_{z,i}^{}b_{z,i}^{}} } 
\end{array}} \right) \hfill \\ 
\end{gathered} 
\end{small}
\end{equation}
where it is noticed that
\begin{equation}
{{\mathbf{B}}^T} = \left( {\begin{array}{*{20}{c}}
  {{H_{x1}}}&{{H_{x2}}}&{{H_{x3}}} \\ 
  {{H_{y1}}}&{{H_{y2}}}&{{H_{y3}}} \\ 
  {{H_{z1}}}&{{H_{z2}}}&{{H_{z3}}} 
\end{array}} \right) = \left( {{\mathbf{h}}_x^T,{\mathbf{h}}_y^T,{\mathbf{h}}_z^T} \right)
\end{equation}
\end{enumerate}

In principle, $\bf{P}$, $\bf{G}$, $\bf{K}$ and $\bf{W}$ have similar formulation of their characteristic polynomials:
\begin{equation}
{\lambda ^4} + {\tau _1}{\lambda ^2} + {\tau _2}\lambda  + {\tau _3} = 0
\end{equation}
where $\tau_{\cdots}$ are coefficients. This is because this form of quartic equation of $\lambda$ is equivalent to the general form \cite{Shmakov2011}
\begin{equation}
{\lambda ^4} + {{\tilde \tau }_1}{\lambda ^3} + {{\tilde \tau }_2}{\lambda ^2} + {{\tilde \tau }_3}\lambda  + {{\tilde \tau }_4} = 0
\end{equation}
where
\begin{equation}
\begin{gathered}
  {\tau _1} = \tilde \tau _2^{} - \frac{3}{8}\tilde \tau _1^2 \hfill \\
  {\tau _2} = \tilde \tau _3^{} - \frac{{\tilde \tau _1^{}\tilde \tau _2^{}}}{2} + \frac{{\tilde \tau _1^3}}{8} \hfill \\
  {\tau _3} = \tilde \tau _4^{} - \frac{{\tilde \tau _1^{}\tilde \tau _3^{}}}{4} + \frac{{\tilde \tau _1^2\tilde \tau _2^{}}}{{16}} - \frac{{3\tilde \tau _1^4}}{{256}} \hfill \\ 
\end{gathered} 
\end{equation}
It can also be seen that $\bf{P}$ and $\bf{K}$ have very similar structure. For $\bf{D} = B$, the sorted eigenvalues of $\bf{P}$ are the same with that of $\bf{K}$.

\setcounter{equation}{35}
\begin{figure*}[hb]
\begin{equation}\label{H_tildes}
\begin{gathered}
  {{{\mathcal{\tilde H}}}_3} =  \hfill \\
  \left( {\begin{array}{*{20}{c}}
  { - 2{H_{x2}}{H_{y1}} + 2{H_{x1}}{H_{y2}}}&{ - 2{H_{x2}}{H_{z1}} + 2{H_{x1}}{H_{z2}}}&{ - 2{H_{y2}}{H_{z1}} + 2{H_{y1}}{H_{z2}}}&0 \\ 
  { - 2{H_{x2}}{H_{z1}} + 2{H_{x1}}{H_{z2}}}&{2{H_{x2}}{H_{y1}} - 2{H_{x1}}{H_{y2}}}&0&{ - 2{H_{y2}}{H_{z1}} + 2{H_{y1}}{H_{z2}}} \\ 
  { - 2{H_{y2}}{H_{z1}} + 2{H_{y1}}{H_{z2}}}&0&{2{H_{x2}}{H_{y1}} - 2{H_{x1}}{H_{y2}}}&{2{H_{x2}}{H_{z1}} - 2{H_{x1}}{H_{z2}}} \\ 
  0&{ - 2{H_{y2}}{H_{z1}} + 2{H_{y1}}{H_{z2}}}&{2{H_{x2}}{H_{z1}} - 2{H_{x1}}{H_{z2}}}&{ - 2{H_{x2}}{H_{y1}} + 2{H_{x1}}{H_{y2}}} 
\end{array}} \right) \hfill \\
  {{{\mathcal{\tilde H}}}_2} =  \hfill \\
  \left( {\begin{array}{*{20}{c}}
  { - 2{H_{x3}}{H_{z1}} + 2{H_{x1}}{H_{z3}}}&{2{H_{x3}}{H_{y1}} - 2{H_{x1}}{H_{y3}}}&0&{ - 2{H_{y3}}{H_{z1}} + 2{H_{y1}}{H_{z3}}} \\ 
  {2{H_{x3}}{H_{y1}} - 2{H_{x1}}{H_{y3}}}&{2{H_{x3}}{H_{z1}} - 2{H_{x1}}{H_{z3}}}&{2{H_{y3}}{H_{z1}} - 2{H_{y1}}{H_{z3}}}&0 \\ 
  0&{2{H_{y3}}{H_{z1}} - 2{H_{y1}}{H_{z3}}}&{ - 2{H_{x3}}{H_{z1}} + 2{H_{x1}}{H_{z3}}}&{2{H_{x3}}{H_{y1}} - 2{H_{x1}}{H_{y3}}} \\ 
  { - 2{H_{y3}}{H_{z1}} + 2{H_{y1}}{H_{z3}}}&0&{2{H_{x3}}{H_{y1}} - 2{H_{x1}}{H_{y3}}}&{2{H_{x3}}{H_{z1}} - 2{H_{x1}}{H_{z3}}} 
\end{array}} \right) \hfill \\
  {{{\mathcal{\tilde H}}}_1} =  \hfill \\
  \left( {\begin{array}{*{20}{c}}
  { - 2{H_{y3}}{H_{z2}} + 2{H_{y2}}{H_{z3}}}&0&{2{H_{x3}}{H_{y2}} - 2{H_{x2}}{H_{y3}}}&{2{H_{x3}}{H_{z2}} - 2{H_{x2}}{H_{z3}}} \\ 
  0&{ - 2{H_{y3}}{H_{z2}} + 2{H_{y2}}{H_{z3}}}&{2{H_{x3}}{H_{z2}} - 2{H_{x2}}{H_{z3}}}&{ - 2{H_{x3}}{H_{y2}} + 2{H_{x2}}{H_{y3}}} \\ 
  {2{H_{x3}}{H_{y2}} - 2{H_{x2}}{H_{y3}}}&{2{H_{x3}}{H_{z2}} - 2{H_{x2}}{H_{z3}}}&{2{H_{y3}}{H_{z2}} - 2{H_{y2}}{H_{z3}}}&0 \\ 
  {2{H_{x3}}{H_{z2}} - 2{H_{x2}}{H_{z3}}}&{ - 2{H_{x3}}{H_{y2}} + 2{H_{x2}}{H_{y3}}}&0&{2{H_{y3}}{H_{z2}} - 2{H_{y2}}{H_{z3}}} 
\end{array}} \right) \hfill \\ 
\end{gathered} 
\end{equation}
\end{figure*}

For Davenport q-method, the quaternion is defined to be vector-wise while for other EIG solvers, the quaternion is scalar-wise, such that
\setcounter{equation}{19}
\begin{equation}\label{quat}
\begin{gathered}
  {{\mathbf{q}}_{\mathbf{K}}} = \left( {\begin{gathered}
  {{\mathbf{n}}\sin \frac{\theta }{2}} \\ 
  {\cos \frac{\theta }{2}} 
\end{gathered}} \right) = {\left( {{q_1},{q_2},{q_3},{q_0}} \right)^T} \hfill \\
  {{\mathbf{q}}_{\mathbf{W}}} = \left( {\begin{gathered}
  {\cos \frac{\theta }{2}} \\ 
  {{\mathbf{n}}\sin \frac{\theta }{2}} 
\end{gathered}} \right) = {\left( {{q_0},{q_1},{q_2},{q_3}} \right)^T} \hfill \\ 
\end{gathered} 
\end{equation}
where $\bf{n}$ is the immediate rotation axis (also known as the unit rotation vector) and $\theta$ denotes the angle about this axis. We have also proved recently that $\bf{K}$ and $\bf{W}$ have the same eigenvalues \cite{Wu2018ase}. That is to say the EIG processes of the matrices $\bf{P, K, W}$ are essentially identical.

\subsection{Analytical Eigenvector Solution}
We first decompose $\bf{W}$ by
\begin{equation}
{\bf{W}} = {{\cal{H}}_1}({\bf{h}}_{\bf{x}}) + {{\cal{H}}_2}({\bf{h}}_{\bf{y}}) + {{\cal{H}}_3}({\bf{h}}_{\bf{z}})
\end{equation}
where
\begin{equation}
\begin{array}{l}
{{\cal{H}}_1}({\bf{h}}_{\bf{x}}) = \left( {\begin{array}{*{20}{c}}
{{H_{x1}}}&0&{ - {H_{z1}}}&{{H_{y1}}}\\
0&{{H_{x1}}}&{{H_{y1}}}&{{H_{z1}}}\\
{ - {H_{z1}}}&{{H_{y1}}}&{ - {H_{x1}}}&0\\
{{H_{y1}}}&{{H_{z1}}}&0&{ - {H_{x1}}}
\end{array}} \right)\\
{{\cal{H}}_2}({\bf{h}}_{\bf{y}}) = \left( {\begin{array}{*{20}{c}}
{{H_{y2}}}&{{H_{z2}}}&0&{ - {H_{x2}}}\\
{{H_{z2}}}&{ - {H_{y2}}}&{{H_{x2}}}&0\\
0&{{H_{x2}}}&{{H_{y2}}}&{{H_{z2}}}\\
{ - {H_{x2}}}&0&{{H_{z2}}}&{ - {H_{y2}}}
\end{array}} \right)\\
{{\cal{H}}_3}({\bf{h}}_{\bf{z}}) = \left( {\begin{array}{*{20}{c}}
{{H_{z3}}}&{ - {H_{y3}}}&{{H_{x3}}}&0\\
{ - {H_{y3}}}&{ - {H_{z3}}}&0&{{H_{x3}}}\\
{{H_{x3}}}&0&{ - {H_{z3}}}&{{H_{y3}}}\\
0&{{H_{x3}}}&{{H_{y3}}}&{{H_{z3}}}
\end{array}} \right)
\end{array}
\end{equation}
An interesting fact about ${\cal{H}}_1$, ${\cal{H}}_2$ and ${\cal{H}}_3$ is that
\begin{equation}
\begin{gathered}
  {{\cal{H}}_2}({\mathbf{x}}) = {\cal{M}}{{\cal{H}}_1}({\mathbf{x}}) \hfill \\
  {{\cal{H}}_3}({\mathbf{x}}) = {\cal{N}}{{\cal{H}}_1}({\mathbf{x}}) \hfill \\ 
\end{gathered} 
\end{equation}
for an arbitrary 3D vector $\bf{x}$, where
\begin{equation}
{\cal{M}} = \left( {\begin{array}{*{20}{c}}
  0&0&0&1 \\ 
  0&0&{ - 1}&0 \\ 
  0&1&0&0 \\ 
  { - 1}&0&0&0 
\end{array}} \right),{\cal{N}} = \left( {\begin{array}{*{20}{c}}
  0&0&{ - 1}&0 \\ 
  0&0&0&{ - 1} \\ 
  1&0&0&0 \\ 
  0&1&0&0 
\end{array}} \right)
\end{equation}
$\cal{M}$ and $\cal{N}$ satisfy
\begin{equation}
\begin{gathered}
  {{\cal{M}}^2} = {{\cal{N}}^2} =  - {\mathbf{I}} \hfill \\
  {\cal{M}}{{\cal{M}}^T} = {{\cal{M}}^T}{\cal{M}} = {\mathbf{I}} \hfill \\
  {\cal{N}}{{\cal{N}}^T} = {{\cal{N}}^T}{\cal{N}} = {\mathbf{I}} \hfill \\ 
\end{gathered} 
\end{equation}
Also for ${\cal{H}}_1$, ${\cal{H}}_2$, ${\cal{H}}_3$, one can observe that
\begin{equation}
{\cal{H}}_1^T({\mathbf{x}}) = {\cal{H}}_1^{}({\mathbf{x}}),{\cal{H}}_2^T({\mathbf{x}}) = {\cal{H}}_2^{}({\mathbf{x}}),{\cal{H}}_3^T({\mathbf{x}}) = {\cal{H}}_3^{}({\mathbf{x}})
\end{equation}
and
\begin{equation}
{\cal{H}}_1^2({\mathbf{x}}) = {\left\| {\mathbf{x}} \right\|^2}{\mathbf{I}}
\end{equation}
which leads to 
\begin{equation}\label{H_norm}
{\cal{H}}_1^2({\mathbf{x}}) = {\cal{H}}_2^2({\mathbf{x}}) = {\cal{H}}_3^2({\mathbf{x}}) = {\left\| {\mathbf{x}} \right\|^2}{\mathbf{I}}
\end{equation}
Let us write out the eigenvalue decomposition of $\bf{W}$
\begin{equation}
{\mathbf{W}} = {{\mathbf{Q}}_{\mathbf{W}}}{{\bf{\Sigma}} _{\mathbf{W}}}{\mathbf{Q}}_{\mathbf{W}}^{ - 1}
\end{equation}
in which ${{\bf{\Sigma}} _{\mathbf{W}}}$ contains two positive and two negative eigenvalues of $\bf{W}$ as $\bf{W}$ is actually indefinite. For the setting of attitude-only estimation from normalized vector pairs, the maximum eigenvalue would not exceed 1. Therefore $\bf{W+I}$ is a positive semidefinite matrix. The eigenvector belonging to ${\lambda _{\max ,{\mathbf{W}}}}$ also belongs to the maximum eigenvalue of $\bf{W+I}$. In numerical analysis, the eigenvector of a matrix's eigenvalue with maximum absolute value can be computed via the power method. The power method is based on the matrix power of
\begin{equation}
{{\mathbf{W}}^n} = {{\mathbf{Q}}_{\mathbf{W}}}{\bf{\Sigma}} _{\mathbf{W}}^n{\mathbf{Q}}_{\mathbf{W}}^{ - 1}
\end{equation}
As $n$ increases, smaller eigenvalues gradually vanish in the final results. For $\bf{W}$, the eigenvalue with maximum absolute value may not always be its positive largest eigenvalue. However, we can perform matrix power on $\bf{W+I}$. The second order power of $\bf{W+I}$ can be computed by
\begin{equation}
{\left( {{\mathbf{W}} + {\mathbf{I}}} \right)^2} = {{\mathbf{W}}^2} + 2{\mathbf{W}} + {\mathbf{I}}
\end{equation}
where
\begin{equation}
\begin{gathered}
  {{\mathbf{W}}^2} = {\cal{H}}_1^2\left( {{{\mathbf{h}}_x}} \right) + {\cal{H}}_2^2\left( {{{\mathbf{h}}_y}} \right) + {\cal{H}}_3^2\left( {{{\mathbf{h}}_z}} \right) +  \hfill \\
  \begin{array}{*{20}{c}}
  {}&{}&{}&{} 
\end{array}{{\cal{H}}_1}\left( {{{\mathbf{h}}_x}} \right){{\cal{H}}_2}\left( {{{\mathbf{h}}_y}} \right) + {{\cal{H}}_2}\left( {{{\mathbf{h}}_y}} \right){{\cal{H}}_1}\left( {{{\mathbf{h}}_x}} \right) +  \hfill \\
  \begin{array}{*{20}{c}}
  {}&{}&{}&{} 
\end{array}{{\cal{H}}_1}\left( {{{\mathbf{h}}_x}} \right){{\cal{H}}_3}\left( {{{\mathbf{h}}_z}} \right) + {{\cal{H}}_3}\left( {{{\mathbf{h}}_z}} \right){{\cal{H}}_1}\left( {{{\mathbf{h}}_x}} \right) +  \hfill \\
  \begin{array}{*{20}{c}}
  {}&{}&{}&{} 
\end{array}{{\cal{H}}_2}\left( {{{\mathbf{h}}_y}} \right){{\cal{H}}_3}\left( {{{\mathbf{h}}_z}} \right) + {{\cal{H}}_3}\left( {{{\mathbf{h}}_z}} \right){{\cal{H}}_2}\left( {{{\mathbf{h}}_y}} \right) \hfill \\ 
\end{gathered} 
\end{equation}
Using (\ref{H_norm}), we have
\begin{equation}
\begin{array}{*{20}{l}}
  {{\mathcal{H}}_1^2 ({\bf{h}}_{{x}}) = \left( {H_{x1}^2 + H_{y1}^2 + H_{z1}^2} \right){\mathbf{I}}} \\ 
  {{\mathcal{H}}_2^2 ({\bf{h}}_{{y}}) = \left( {H_{x2}^2 + H_{y2}^2 + H_{z2}^2} \right){\mathbf{I}}} \\ 
  {{\mathcal{H}}_3^2 ({\bf{h}}_{{z}}) = \left( {H_{x3}^2 + H_{y3}^2 + H_{z3}^2} \right){\mathbf{I}}} 
\end{array}
\end{equation}
One can rearrange the components of ${\bf{W}}^2$ to the following form
\begin{equation}
\begin{gathered}
  {{{\mathcal{\tilde H}}}_3} = {{\mathcal{H}}_1}({\bf{h}}_{{x}}){{\mathcal{H}}_2}({\bf{h}}_{{y}}) + {{\mathcal{H}}_2}({\bf{h}}_{{y}}){{\mathcal{H}}_1}({\bf{h}}_{{x}}) \hfill \\
  {{{\mathcal{\tilde H}}}_2} = {{\mathcal{H}}_1}({\bf{h}}_{{x}}){{\mathcal{H}}_3}({\bf{h}}_{{z}}) + {{\mathcal{H}}_3}({\bf{h}}_{{z}}){{\mathcal{H}}_1}({\bf{h}}_{{x}}) \hfill \\
  {{{\mathcal{\tilde H}}}_1} = {{\mathcal{H}}_2}({\bf{h}}_{{y}}){{\mathcal{H}}_3}({\bf{h}}_{{z}}) + {{\mathcal{H}}_3}({\bf{h}}_{{z}}){{\mathcal{H}}_2}({\bf{h}}_{{y}}) \hfill \\ 
\end{gathered} 
\end{equation}
where ${{{\mathcal{\tilde H}}}_3},{{{\mathcal{\tilde H}}}_2},{{{\mathcal{\tilde H}}}_1}$ are shown in (\ref{H_tildes}). We can see that ${{{\mathcal{\tilde H}}}_3},{{{\mathcal{\tilde H}}}_2},{{{\mathcal{\tilde H}}}_1}$ own the same forms of ${{{\mathcal{H}}}_3},{{{\mathcal{H}}}_2},{{{\mathcal{H}}}_1}$ respectively generating the following mapping
\begin{equation}
\begin{gathered}
  {{\cal{T}}_1}:\left\{ {\begin{array}{*{20}{c}}
  {{{\tilde H}_{x3}} \leftarrow  - 2\left( {{H_{y2}}{H_{z1}} - {H_{y1}}{H_{z2}}} \right)} \\ 
  {{{\tilde H}_{y3}} \leftarrow 2\left( {{H_{x2}}{H_{z1}} - {H_{x1}}{H_{z2}}} \right)} \\ 
  {{{\tilde H}_{z3}} \leftarrow  - 2\left( {{H_{x2}}{H_{y1}} - {H_{x1}}{H_{y2}}} \right)} 
\end{array}} \right. \hfill \\
  {{\cal{T}}_2}:\left\{ {\begin{array}{*{20}{c}}
  {{{\tilde H}_{x2}} \leftarrow 2\left( {{H_{y3}}{H_{z1}} - {H_{y1}}{H_{z3}}} \right)} \\ 
  {{{\tilde H}_{y2}} \leftarrow  - 2\left( {{H_{x3}}{H_{z1}} - {H_{x1}}{H_{z3}}} \right)} \\ 
  {{{\tilde H}_{z2}} \leftarrow 2\left( {{H_{x3}}{H_{y1}} - {H_{x1}}{H_{y3}}} \right)} 
\end{array}} \right. \hfill \\
  {{\cal{T}}_3}:\left\{ {\begin{array}{*{20}{c}}
  {{{\tilde H}_{x1}} \leftarrow  - 2\left( {{H_{y3}}{H_{z2}} - {H_{y2}}{H_{z3}}} \right)} \\ 
  {{{\tilde H}_{y1}} \leftarrow 2\left( {{H_{x3}}{H_{z2}} - {H_{x2}}{H_{z3}}} \right)} \\ 
  {{{\tilde H}_{z1}} \leftarrow  - 2\left( {{H_{x3}}{H_{y2}} - {H_{x2}}{H_{y3}}} \right)} 
\end{array}} \right. \hfill \\ 
\end{gathered} 
\end{equation}
where ${{\tilde H}_{\cdots} }$ denote the entries of ${{{\mathcal{\tilde H}}}_3},{{{\mathcal{\tilde H}}}_2},{{{\mathcal{\tilde H}}}_1}$. The above mappings are in the following cross product evolution
\setcounter{equation}{36}
\begin{equation}
\begin{gathered}
  {{\cal{T}}_1}:{{{\mathbf{\tilde h}}}_z} = \left( {{{\tilde H}_{x1}},{{\tilde H}_{y1}},{{\tilde H}_{z1}}} \right) \leftarrow 2{{\mathbf{h}}_x} \times {{\mathbf{h}}_y} \hfill \\
  {{\cal{T}}_2}:{{{\mathbf{\tilde h}}}_y} = \left( {{{\tilde H}_{x2}},{{\tilde H}_{y2}},{{\tilde H}_{z2}}} \right) \leftarrow 2{{\mathbf{h}}_z} \times {{\mathbf{h}}_x} \hfill \\
  {{\cal{T}}_3}:{{{\mathbf{\tilde h}}}_x} = \left( {{{\tilde H}_{x3}},{{\tilde H}_{y3}},{{\tilde H}_{z3}}} \right) \leftarrow 2{{\mathbf{h}}_y} \times {{\mathbf{h}}_z} \hfill \\ 
\end{gathered} 
\end{equation}
Meanwhile, notice that ${{{\mathcal{\tilde H}}}_3},{{{\mathcal{\tilde H}}}_2},{{{\mathcal{\tilde H}}}_1}$ are additive i.e.
\begin{equation}
\begin{gathered}
  {{\cal{H}}_1}({\mathbf{x}} + {\mathbf{y}}) = {{\cal{H}}_1}({\mathbf{x}}) + {{\cal{H}}_1}({\mathbf{y}}) \hfill \\
  {{\cal{H}}_2}({\mathbf{x}} + {\mathbf{y}}) = {{\cal{H}}_2}({\mathbf{x}}) + {{\cal{H}}_2}({\mathbf{y}}) \hfill \\
  {{\cal{H}}_3}({\mathbf{x}} + {\mathbf{y}}) = {{\cal{H}}_3}({\mathbf{x}}) + {{\cal{H}}_3}({\mathbf{y}}) \hfill \\ 
\end{gathered} 
\end{equation}
for two arbitrary 3D vectors $\bf{x,y}$. Then $({\bf{W+I}})^2$ can be rewritten into
\begin{equation}
\begin{gathered}
  {\left( {{\mathbf{W}} + {\mathbf{I}}} \right)^2} = \left( {{{\left\| {{{\mathbf{h}}_x}} \right\|}^2} + {{\left\| {{{\mathbf{h}}_y}} \right\|}^2} + {{\left\| {{{\mathbf{h}}_z}} \right\|}^2}} + 1 \right){\mathbf{I}} +  \hfill \\
  \begin{array}{*{20}{c}}
  {}&{} 
\end{array}2{{\cal{H}}_1}\left( {{{\mathbf{h}}_x} + {{{\mathbf{\tilde h}}}_x}} \right) + 2{{\cal{H}}_2}\left( {{{\mathbf{h}}_y} + {{{\mathbf{\tilde h}}}_y}} \right) + 2{{\cal{H}}_3}\left( {{{\mathbf{h}}_z} + {{{\mathbf{\tilde h}}}_z}} \right) \hfill \\ 
\end{gathered} 
\end{equation}
Letting 
\begin{equation}
{\mathbf{F}} = \frac{1}{{{{\left\| {{{\mathbf{h}}_x}} \right\|}^2} + {{\left\| {{{\mathbf{h}}_y}} \right\|}^2} + {{\left\| {{{\mathbf{h}}_z}} \right\|}^2} + 1}}{\left( {{\mathbf{W}} + {\mathbf{I}}} \right)^2}
\end{equation}
, we can see that $\bf{F}$ is in the form of $\bf{W+I}$ such that ${\mathbf{F}} = {\mathbf{\tilde W}} + {\mathbf{I}}$ where
\begin{equation}
{\mathbf{\tilde W}} = \frac{2}{{{{\left\| {{{\mathbf{h}}_x}} \right\|}^2} + {{\left\| {{{\mathbf{h}}_y}} \right\|}^2} + {{\left\| {{{\mathbf{h}}_z}} \right\|}^2} + 1}}\left[ \begin{gathered}
  {{\cal{H}}_1}\left( {{{\mathbf{h}}_x} + {{{\mathbf{\tilde h}}}_x}} \right) +  \hfill \\
  {{\cal{H}}_2}\left( {{{\mathbf{h}}_y} + {{{\mathbf{\tilde h}}}_y}} \right) +  \hfill \\
  {{\cal{H}}_3}\left( {{{\mathbf{h}}_z} + {{{\mathbf{\tilde h}}}_z}} \right) \hfill \\ 
\end{gathered}  \right]
\end{equation}
That is to say, the power of $\bf{W+I}$ can be indirectly computed via the power of $\bf{F}$ and so on. Finally, the following iteration is designed to achieve the recursive propagation
\begin{equation}\label{iteration}
\begin{gathered}
  {{\mathbf{h}}_{x,k}} = {\rho _{k - 1}}\left( {{{\mathbf{h}}_{x,k - 1}} + {{\mathbf{h}}_{y,k - 1}} \times {{\mathbf{h}}_{z,k - 1}}} \right) \hfill \\
  {{\mathbf{h}}_{y,k}} = {\rho _{k - 1}}\left( {{{\mathbf{h}}_{y,k - 1}} + {{\mathbf{h}}_{z,k - 1}} \times {{\mathbf{h}}_{x,k - 1}}} \right) \hfill \\
  {{\mathbf{h}}_{z,k}} = {\rho _{k - 1}}\left( {{{\mathbf{h}}_{z,k - 1}} + {{\mathbf{h}}_{x,k - 1}} \times {{\mathbf{h}}_{y,k - 1}}} \right) \hfill \\
  {\rho _{k - 1}} = \frac{2}{{{{\left\| {{{\mathbf{h}}_{x,k - 1}}} \right\|}^2} + {{\left\| {{{\mathbf{h}}_{y,k - 1}}} \right\|}^2} + {{\left\| {{{\mathbf{h}}_{z,k - 1}}} \right\|}^2} + 1}} \hfill \\ 
\end{gathered} 
\end{equation}
where $k$ is the iteration index for $k=1,2,\cdots$; $\rho$ is the quasi-normalization factor and the initial vectors are
\begin{equation}
{{\mathbf{h}}_{x,1}} = {{\mathbf{h}}_x},{{\mathbf{h}}_{y,1}} = {{\mathbf{h}}_y},{{\mathbf{h}}_{z,1}} = {{\mathbf{h}}_z}
\end{equation}
Assume that we conduct $k$ iterations for the system (\ref{iteration}), then the matrix bases of 
\begin{equation}
\gamma {\left( {{\mathbf{W}} + {\mathbf{I}}} \right)^{{2^j}}}
\end{equation}
where $\gamma$ is a constant formed by $\rho_{k}, k=1,2,\cdots$, can be computed within short time. The final quaternion solution to this eigenvalue problem can be taken by the following multiplying
\begin{equation}
{\mathbf{q}} = \frac{{\gamma {{\left( {{\mathbf{W}} + {\mathbf{I}}} \right)}^{{2^j}}}{{\left( {1,0,0,0} \right)}^T}}}{{\left\| {\gamma {{\left( {{\mathbf{W}} + {\mathbf{I}}} \right)}^{{2^j}}}{{\left( {1,0,0,0} \right)}^T}} \right\|}}
\end{equation}
whose expression is detailed as follows
\begin{equation}\label{quat}
\begin{gathered}
  {\mathbf{q}} = \frac{1}{{\sqrt{ \begin{gathered}
  {\left( {{H_{x1}} + {H_{y2}} + {H_{z3}}} \right)^2} + {\left( { - {H_{y3}} + {H_{z2}}} \right)^2} +  \hfill \\
  {\left( { - {H_{z1}} + {H_{x3}}} \right)^2} + {\left( { - {H_{x2}} + {H_{y1}}} \right)^2} \hfill \\ 
\end{gathered}}  }} \times  \hfill \\
  \begin{array}{*{20}{c}}
  {}&{}&{}&{} &{}&{}&{}&{} 
\end{array}\left( {\begin{array}{*{20}{c}}
  {{H_{x1}} + {H_{y2}} + {H_{z3}}} \\ 
  { - {H_{y3}} + {H_{z2}}} \\ 
  { - {H_{z1}} + {H_{x3}}} \\ 
  { - {H_{x2}} + {H_{y1}}} 
\end{array}} \right) \hfill \\ 
\end{gathered} 
\end{equation}
The advantage of the proposed method is that it does not have to compute the $4\times4$ symmetric matrix and has no need to calculate the coefficients of its characteristic polynomial and the roots. The direct design makes the proposed method faster than existing representative ones in real tests.

\subsection{Convergence}
\indent The convergence of the above iterative system (\ref{iteration}) can be characterized with
%\begin{equation}
%\begin{gathered}
%  {{\mathbf{h}}_{x,k}} = {\rho _{k - 1}}\left( {{{\mathbf{h}}_{x,k - 1}} + {{\mathbf{h}}_{y,k - 1}} \times {{\mathbf{h}}_{z,k - 1}}} \right) \hfill \\
%   \Rightarrow (1 - {\rho _{k - 1}}){{\mathbf{h}}_{x,k}} =  - {\rho _{k - 1}}\left( {{{\mathbf{h}}_{x,k}} - {{\mathbf{h}}_{x,k - 1}}} \right) \hfill \\
%  \begin{array}{*{20}{c}}
%  {}&{}&{}&{}&{}&{}&{}&{}&{}
%\end{array} + {\rho _{k - 1}}\left( {{{\mathbf{h}}_{y,k - 1}} \times {{\mathbf{h}}_{z,k - 1}}} \right) \hfill \\
%   \Rightarrow {{\mathbf{h}}_{x,k}} =  - \frac{{{\rho _{k - 1}}}}{{1 - {\rho _{k - 1}}}}\left( {\Delta {{\mathbf{h}}_{x,k}} - {{\mathbf{h}}_{y,k - 1}} \times {{\mathbf{h}}_{z,k - 1}}} \right) \hfill \\ 
%\end{gathered} 
%\end{equation}
%where $\Delta {{\mathbf{h}}_{x,k}} = {{\mathbf{h}}_{x,k}} - {{\mathbf{h}}_{x,k - 1}}$ is the backward differentiation. 
the following constraint 
\begin{equation}
\left\| {{\mathbf{x}} \times {\mathbf{y}}} \right\| = \left\| {\mathbf{x}} \right\|\left\| {\mathbf{y}} \right\|\left| {\sin \left\langle {{\mathbf{x}},{\mathbf{y}}} \right\rangle } \right| \le \min \left( {\left\| {\mathbf{x}} \right\|,\left\| {\mathbf{y}} \right\|} \right)
\end{equation}
Therefore the following inequalities can be obtained
\begin{equation}
\begin{gathered}
  {\left\| {{{\mathbf{h}}_{x,k}}} \right\|^2} \hfill \\
   = \rho _{k - 1}^2{\left\| {{{\mathbf{h}}_{x,k - 1}} + {{\mathbf{h}}_{y,k - 1}} \times {{\mathbf{h}}_{z,k - 1}}} \right\|^2} \hfill \\
   \leqslant \rho _{k - 1}^2\left( {{{\left\| {{{\mathbf{h}}_{x,k - 1}}} \right\|}^2} + {{\left\| {{{\mathbf{h}}_{y,k - 1}} \times {{\mathbf{h}}_{z,k - 1}}} \right\|}^2}} \right) \hfill \\
   \leqslant \rho _{k - 1}^2\left[ {{{\left\| {{{\mathbf{h}}_{x,k - 1}}} \right\|}^2} + \frac{1}{2}\left( {{{\left\| {{{\mathbf{h}}_{y,k - 1}}} \right\|}^2} + {{\left\| {{{\mathbf{h}}_{z,k - 1}}} \right\|}^2}} \right)} \right] \hfill \\
   = \rho _{k - 1}^2\left[ {{{\left\| {{{\mathbf{h}}_{x,k - 1}}} \right\|}^2} + \frac{1}{2}\left( {{{\left\| {{{\mathbf{h}}_{y,k - 1}}} \right\|}^2} + {{\left\| {{{\mathbf{h}}_{z,k - 1}}} \right\|}^2}} \right)} \right] \hfill \\
   = \frac{1}{2}\rho _{k - 1}^2\left[ {2{{\left\| {{{\mathbf{h}}_{x,k - 1}}} \right\|}^2} + {{\left\| {{{\mathbf{h}}_{y,k - 1}}} \right\|}^2} + {{\left\| {{{\mathbf{h}}_{z,k - 1}}} \right\|}^2}} \right] \hfill \\
   = \frac{1}{2}\rho _{k - 1}^{}\frac{{2{{\left\| {{{\mathbf{h}}_{x,k - 1}}} \right\|}^2} + {{\left\| {{{\mathbf{h}}_{y,k - 1}}} \right\|}^2} + {{\left\| {{{\mathbf{h}}_{z,k - 1}}} \right\|}^2}}}{{{{\left\| {{{\mathbf{h}}_{x,k - 1}}} \right\|}^2} + {{\left\| {{{\mathbf{h}}_{y,k - 1}}} \right\|}^2} + {{\left\| {{{\mathbf{h}}_{z,k - 1}}} \right\|}^2} + 1}} \hfill \\
   \leqslant \frac{1}{2}\rho _{k - 1}^{}\frac{{{{\left\| {{{\mathbf{h}}_{x,k - 1}}} \right\|}^2} + {{\left\| {{{\mathbf{h}}_{y,k - 1}}} \right\|}^2} + {{\left\| {{{\mathbf{h}}_{z,k - 1}}} \right\|}^2} + 1}}{{{{\left\| {{{\mathbf{h}}_{x,k - 1}}} \right\|}^2} + {{\left\| {{{\mathbf{h}}_{y,k - 1}}} \right\|}^2} + {{\left\| {{{\mathbf{h}}_{z,k - 1}}} \right\|}^2} + 1}} \hfill \\
   = \frac{1}{2}\rho _{k - 1}^{} \leqslant \frac{1}{2}\rho _{k - 1}^{}{\left\| {{{\mathbf{h}}_{x,k - 1}}} \right\|^2} \hfill \\
   < {\left\| {{{\mathbf{h}}_{x,k - 1}}} \right\|^2} \hfill \\ 
\end{gathered} 
\end{equation}
for $\left\| {{{\mathbf{h}}_{x,k - 1}}} \right\| \geqslant 1$. And for $\left\| {{{\mathbf{h}}_{x,k - 1}}} \right\| \leqslant 1$, we have ${\left\| {{{\mathbf{h}}_{x,k}}} \right\|^2} > {\left\| {{{\mathbf{h}}_{x,k - 1}}} \right\|^2}$. Then invoking the famous squeeze theorem and applying it to other two vectors, we arrive at $\mathop {\lim }\limits_{k \to  + \infty } {\left\| {{{\mathbf{h}}_{x,k}}} \right\|^2} = \mathop {\lim }\limits_{k \to  + \infty } {\left\| {{{\mathbf{h}}_{y,k}}} \right\|^2} = \mathop {\lim }\limits_{k \to  + \infty } {\left\| {{{\mathbf{h}}_{z,k}}} \right\|^2} = 1$.
At this point, the proposed algorithm is definitely convergent and stable. The final procedure of the proposed quaternion solution to rigid 3D registration is listed in Algorithm \ref{algorithm:FA3R}.

\subsection{Steady-State Evolution and Optimal Rotation Matrix}\label{sec:evo}
As described in last sub-section, when $k \to  + \infty $, ${{\mathbf{h}}_x},{{\mathbf{h}}_y},{{\mathbf{h}}_z}$ converge to constant 1. The steady-state system can be written as follows
\begin{equation}
\begin{gathered}
  {{\mathbf{h}}_{x,\infty }} = {\rho _\infty }\left( {{{\mathbf{h}}_{x,\infty }} + {{\mathbf{h}}_{y,\infty }} \times {{\mathbf{h}}_{z,\infty }}} \right) \hfill \\
  {{\mathbf{h}}_{y,\infty }} = {\rho _\infty }\left( {{{\mathbf{h}}_{y,\infty }} + {{\mathbf{h}}_{z,\infty }} \times {{\mathbf{h}}_{x,\infty }}} \right) \hfill \\
  {{\mathbf{h}}_{z,\infty }} = {\rho _\infty }\left( {{{\mathbf{h}}_{z,\infty }} + {{\mathbf{h}}_{x,\infty }} \times {{\mathbf{h}}_{y,\infty }}} \right) \hfill \\ 
\end{gathered} 
\end{equation}

\begin{algorithm}[H] 
\caption{The simple fast analytical vectorial solution to quaternion for the rigid 3D registration problem.} 
{\textbf{Step 1:}} Compute the cross-covariance matrix:
\begin{equation*}
{\mathbf{D}} = \frac{1}{n}\sum\limits_{i = 1}^n {\left( {{\mathbf{r}}_i^{} - {\mathbf{\bar r}}} \right){{\left( {{\mathbf{b}}_i^{} - {\mathbf{\bar b}}} \right)}^T}}
\end{equation*}
{\textbf{Step 2:}} Set the initial iteration index: $k=1$\\
{\textbf{Step 3:}} Prepare the vectors:
\begin{equation*}
\begin{gathered}
  {{\mathbf{h}}_x} = {\left( {{D_{11}},{D_{21}},{D_{31}}} \right)} \hfill \\
  {{\mathbf{h}}_y} = {\left( {{D_{12}},{D_{22}},{D_{32}}} \right)} \hfill \\
  {{\mathbf{h}}_z} = {\left( {{D_{13}},{D_{23}},{D_{33}}} \right)} \hfill \\
  {{\mathbf{h}}_{x,1}} = {{\mathbf{h}}_x},{{\mathbf{h}}_{y,1}} = {{\mathbf{h}}_y},{{\mathbf{h}}_{z,1}} = {{\mathbf{h}}_z} \hfill \\ 
\end{gathered} 
\end{equation*}
{\textbf{Step 4:}} Relative accuracy threshold $= \epsilon$\\ 
{\textbf{Step 5:}}  \textbf{while}  $\begin{array}{*{20}{c}}
  \begin{gathered}
  {\left\| {{{\mathbf{h}}_{x,k}} - {{\mathbf{h}}_{x,k - 1}}} \right\|^2} +  \hfill \\
  {\left\| {{{\mathbf{h}}_{y,k}} - {{\mathbf{h}}_{y,k - 1}}} \right\|^2} +  \hfill \\
  {\left\| {{{\mathbf{h}}_{z,k}} - {{\mathbf{h}}_{z,k - 1}}} \right\|^2} \hfill \\ 
\end{gathered}  
\end{array} < \epsilon $ \textbf{do} 
\begin{enumerate} 
\item $k=k+1$
\item Compute the quasi-normalization factor:
\begin{equation*}
{\rho _{k - 1}} = \frac{2}{{{{\left\| {{{\mathbf{h}}_{x,k - 1}}} \right\|}^2} + {{\left\| {{{\mathbf{h}}_{y,k - 1}}} \right\|}^2} + {{\left\| {{{\mathbf{h}}_{z,k - 1}}} \right\|}^2} + 1}}
\end{equation*}
\item Do vector iterations by
\begin{equation*}
\begin{gathered}
  {{\mathbf{h}}_{x,k}} = {\rho _{k - 1}}\left( {{{\mathbf{h}}_{x,k - 1}} + {{\mathbf{h}}_{y,k - 1}} \times {{\mathbf{h}}_{z,k - 1}}} \right) \hfill \\
  {{\mathbf{h}}_{y,k}} = {\rho _{k - 1}}\left( {{{\mathbf{h}}_{y,k - 1}} + {{\mathbf{h}}_{z,k - 1}} \times {{\mathbf{h}}_{x,k - 1}}} \right) \hfill \\
  {{\mathbf{h}}_{z,k}} = {\rho _{k - 1}}\left( {{{\mathbf{h}}_{z,k - 1}} + {{\mathbf{h}}_{x,k - 1}} \times {{\mathbf{h}}_{y,k - 1}}} \right) \hfill \\ 
\end{gathered} 
\end{equation*}
\end{enumerate} 
{}
\ \ \ \ \ \ \ {\textbf{end while}} \\
{\textbf{Step 6:} Quaternion computation: Using (\ref{quat})}.
\label{algorithm:FA3R} 
\end{algorithm}

\noindent Letting $\varsigma  = \frac{{1 - {\rho _\infty }}}{{{\rho _\infty }}}$, we have
\begin{equation}
\begin{gathered}
  \varsigma {{\mathbf{h}}_{x,\infty }} = {{\mathbf{h}}_{y,\infty }} \times {{\mathbf{h}}_{z,\infty }} \hfill \\
  \varsigma {{\mathbf{h}}_{y,\infty }} = {{\mathbf{h}}_{z,\infty }} \times {{\mathbf{h}}_{x,\infty }} \hfill \\
  \varsigma {{\mathbf{h}}_{z,\infty }} = {{\mathbf{h}}_{x,\infty }} \times {{\mathbf{h}}_{y,\infty }} \hfill \\ 
\end{gathered} 
\end{equation}
That is to say at this time, ${{\mathbf{h}}_x},{{\mathbf{h}}_y},{{\mathbf{h}}_z}$ are orthogonal with each other. The above result also ensures
\begin{equation}
\begin{small}
\begin{gathered}
  \varsigma \left\| {{{\mathbf{h}}_{x,\infty }}} \right\| = \left\| {{{\mathbf{h}}_{y,\infty }}} \right\|\left\| {{{\mathbf{h}}_{z,\infty }}} \right\|\sin \left\langle {{{\mathbf{h}}_{y,\infty }},{{\mathbf{h}}_{z,\infty }}} \right\rangle  = \left\| {{{\mathbf{h}}_{y,\infty }}} \right\|\left\| {{{\mathbf{h}}_{z,\infty }}} \right\| \hfill \\
  \varsigma \left\| {{{\mathbf{h}}_{y,\infty }}} \right\| = \left\| {{{\mathbf{h}}_{z,\infty }}} \right\|\left\| {{{\mathbf{h}}_{x,\infty }}} \right\|\sin \left\langle {{{\mathbf{h}}_{z,\infty }},{{\mathbf{h}}_{x,\infty }}} \right\rangle  = \left\| {{{\mathbf{h}}_{z,\infty }}} \right\|\left\| {{{\mathbf{h}}_{x,\infty }}} \right\| \hfill \\
  \varsigma \left\| {{{\mathbf{h}}_{z,\infty }}} \right\| = \left\| {{{\mathbf{h}}_{x,\infty }}} \right\|\left\| {{{\mathbf{h}}_{y,\infty }}} \right\|\sin \left\langle {{{\mathbf{h}}_{x,\infty }},{{\mathbf{h}}_{y,\infty }}} \right\rangle  = \left\| {{{\mathbf{h}}_{x,\infty }}} \right\|\left\| {{{\mathbf{h}}_{y,\infty }}} \right\| \hfill \\ 
\end{gathered} 
\end{small}
\end{equation}
Then we immediately arrive at
\begin{equation}\label{k}
\left\| {{{\mathbf{h}}_{x,\infty }}} \right\| = \left\| {{{\mathbf{h}}_{y,\infty }}} \right\| = \left\| {{{\mathbf{h}}_{z,\infty }}} \right\| = \varsigma 
\end{equation}
Inserting (\ref{k}) into (\ref{iteration}), it is obtained that
\begin{equation}
{\rho _\infty } = \frac{2}{{3{\varsigma ^2} + 1}}
\end{equation}
By solving
\begin{equation}
\varsigma  = \frac{{1 - {\rho _\infty }}}{{{\rho _\infty }}} = \frac{{3{\varsigma ^2} + 1}}{2} - 1,\varsigma  > 0
\end{equation}
the steady-state $\varsigma$ equals to 1, which leads to ${\rho _\infty } = \frac{1}{2}$. In such circumstance, the matrix 
\begin{equation}
\begin{gathered}
  {{\cal{C}}_1} = \left( {{\mathbf{h}}_{x,\infty }^T,{\mathbf{h}}_{y,\infty }^T,{\mathbf{h}}_{z,\infty }^T} \right) \hfill \\
  {{\cal{C}}_2} = {\cal{C}}_1^T = \left( {\begin{array}{*{20}{c}}
  {{{\mathbf{h}}_{x,\infty }}} \\ 
  {{{\mathbf{h}}_{y,\infty }}} \\ 
  {{{\mathbf{h}}_{z,\infty }}} 
\end{array}} \right) \hfill \\ 
\end{gathered} 
\end{equation}
are orthonormal matrices since ${{\mathbf{h}}_{x,\infty }} \bot {{\mathbf{h}}_{y,\infty }} \bot {{\mathbf{h}}_{z,\infty }}$ and $\left\| {{{\mathbf{h}}_{x,\infty }}} \right\| = \left\| {{{\mathbf{h}}_{y,\infty }}} \right\| = \left\| {{{\mathbf{h}}_{z,\infty }}} \right\| = 1$. The orthogonal matrix bases are constituted by ${\mathbf{h}}_{x,\infty }^T,{\mathbf{h}}_{y,\infty }^T,{\mathbf{h}}_{z,\infty }^T$. The determinants can be computed as follows
\begin{equation}
\begin{gathered}
  \det \left( {{C_1}} \right) = \det \left( {{C_2}} \right) = {H_{x1}}{H_{y2}}{H_{z3}} - {H_{x1}}{H_{y3}}{H_{z3}} -  \hfill \\
  \begin{array}{*{20}{c}}
  {}&{}&{}&{}&{}&{} 
\end{array}{H_{x2}}{H_{y1}}{H_{z3}} + {H_{x2}}{H_{y3}}{H_{z1}} +  \hfill \\
  \begin{array}{*{20}{c}}
  {}&{}&{}&{}&{}&{} 
\end{array}{H_{x3}}{H_{y1}}{H_{z2}} - {H_{x3}}{H_{y2}}{H_{z1}} \hfill \\
   = {{\mathbf{h}}_{x,\infty }}\left( {{\mathbf{h}}_{y,\infty }^T \times {\mathbf{h}}_{z,\infty }^T} \right) \hfill \\
  \begin{array}{*{20}{c}}
  {}&{}&{} 
\end{array} = {{\mathbf{h}}_{y,\infty }}\left( {{\mathbf{h}}_{z,\infty }^T \times {\mathbf{h}}_{x,\infty }^T} \right) \hfill \\
  \begin{array}{*{20}{c}}
  {}&{}&{}&{}&{}&{} 
\end{array} = {{\mathbf{h}}_{z,\infty }}\left( {{\mathbf{h}}_{x,\infty }^T \times {\mathbf{h}}_{z,\infty }^T} \right) \hfill \\
   = 1 \hfill \\ 
\end{gathered} 
\end{equation}
If we rewrite the $\bf{B}$ into ${\mathbf{B}} = \left( {{{\mathbf{v}}_x},{{\mathbf{v}}_y},{{\mathbf{v}}_z}} \right) = \left( {\begin{array}{*{20}{c}}
  {{\mathbf{h}}_x^{}} \\ 
  {{\mathbf{h}}_y^{}} \\ 
  {{\mathbf{h}}_z^{}} 
\end{array}} \right)$ where ${{\mathbf{v}}_x},{{\mathbf{v}}_y},{{\mathbf{v}}_z}$ are column vectors, with iteration in (\ref{iteration}), the following recursion is generated as well
\begin{equation}
\begin{gathered}
  {{\mathbf{v}}_{x,k}} = {\rho _{k - 1}}\left( {{{\mathbf{v}}_{x,k - 1}} + {{\mathbf{v}}_{y,k - 1}} \times {{\mathbf{v}}_{z,k - 1}}} \right) \hfill \\
  {{\mathbf{v}}_{y,k}} = {\rho _{k - 1}}\left( {{{\mathbf{v}}_{y,k - 1}} + {{\mathbf{v}}_{z,k - 1}} \times {{\mathbf{v}}_{x,k - 1}}} \right) \hfill \\
  {{\mathbf{v}}_{z,k}} = {\rho _{k - 1}}\left( {{{\mathbf{v}}_{z,k - 1}} + {{\mathbf{v}}_{x,k - 1}} \times {{\mathbf{v}}_{y,k - 1}}} \right) \hfill \\
  {\rho _{k - 1}} = \frac{2}{{{{\left\| {{{\mathbf{v}}_{x,k - 1}}} \right\|}^2} + {{\left\| {{{\mathbf{v}}_{y,k - 1}}} \right\|}^2} + {{\left\| {{{\mathbf{v}}_{z,k - 1}}} \right\|}^2} + 1}} \hfill \\ 
\end{gathered} 
\end{equation}
because of
\begin{equation}
\begin{gathered}
  {\left\| {{{\mathbf{v}}_{x,k - 1}}} \right\|^2} + {\left\| {{{\mathbf{v}}_{y,k - 1}}} \right\|^2} + {\left\| {{{\mathbf{v}}_{z,k - 1}}} \right\|^2} \hfill \\
  \begin{array}{*{20}{c}}
  {}&{}&{}&{} 
\end{array} = {\left\| {{{\mathbf{h}}_{x,k - 1}}} \right\|^2} + {\left\| {{{\mathbf{h}}_{y,k - 1}}} \right\|^2} + {\left\| {{{\mathbf{h}}_{z,k - 1}}} \right\|^2} \hfill \\ 
\end{gathered} 
\end{equation}
where $k=2,3,\cdots$ is the iteration index for which ${{\mathbf{v}}_{x,1}} = {{\mathbf{v}}_x},{{\mathbf{v}}_{y,1}} = {{\mathbf{v}}_y},{{\mathbf{v}}_{z,1}} = {{\mathbf{v}}_z}$. Then the final results of ${{\mathbf{v}}_{x,\infty }},{{\mathbf{v}}_{y,\infty }},{{\mathbf{v}}_{z,\infty }}$ are orthonormal bases of ${\cal{C}}_1$ and ${\cal{C}}_2$. These factors enable the following identities
\begin{equation}
\begin{gathered}
  {{\cal{C}}_1}{\cal{C}}_1^T = {\cal{C}}_1^T{\cal{C}}_1^{} = {\mathbf{I}},\det \left( {{\cal{C}}_1^{}} \right) = 1 \hfill \\
  {{\cal{C}}_2}{\cal{C}}_2^T = {\cal{C}}_2^T{\cal{C}}_2^{} = {\mathbf{I}},\det \left( {{\cal{C}}_2^{}} \right) = 1 \hfill \\ 
\end{gathered} 
\end{equation}
i.e. ${\cal{C}}_1^{},{\cal{C}}_2^{} \in SO(3)$. As $\bf{D}$ is 
\begin{equation}
{\mathbf{D}} = \left( {{\mathbf{h}}_{x,1}^T,{\mathbf{h}}_{y,1}^T,{\mathbf{h}}_{z,1}^T} \right)
\end{equation}
, the optimal rotation matrix of the optimization (\ref{opt}) should be ${\cal{C}}_1$.

\subsection{Real-World Implementation}
The designed iteration (\ref{iteration}) can directly obtain the attitude matrix on $SO(3)$ and then produce the quaternion. In real-world implementation, the cross product can be easily computed as follows
\begin{equation}
{\mathbf{x}} \times {\mathbf{y}} = {\left( {{x_2}{y_3} - {x_3}{y_2},{x_3}{y_1} - {x_1}{y_3},{x_1}{y_2} - {x_2}{y_1}} \right)^T}
\end{equation}
where ${\mathbf{x}} = {\left( {{x_1},{x_2},{x_3}} \right)^T},{\mathbf{y}} = {\left( {{y_1},{y_2},{y_3}} \right)^T}$ are two arbitrary 3D vectors. This computation step, unlike the SVD or EIG, can be performed using signed integers. For instance, if one would like to obtain results with single-precision accuracy, then by introducing a temporary float-to-integer conversion of $2^{32}$, the results can be computed using 64-bit integers within very short time. This operation makes sure that the algorithm is feasible in real applications due to the existence of noises. For a captured point cloud from a typical 16-line LiDAR, the noise density is normally between $0.01\%$ and $1\%$. In such occasion, the double precision is required and the 128-bit signed integer conversion should be taken. However, for other platforms e.g. visual navigation systems, the noises are usually much larger than that from LiDAR, which is induced by mismatching of feature points or improper calibration of intrinsic matrix of camera and extrinsic parameters between camera and other aided instruments. In such occasion, the single-precision numbers have been already sufficient for description of results. While for SVD and EIG, there are too many numerical operations inside which limits their extensions with large signed integers.

\begin{figure*}[hb]
\centering
\includegraphics[width=0.75\textwidth]{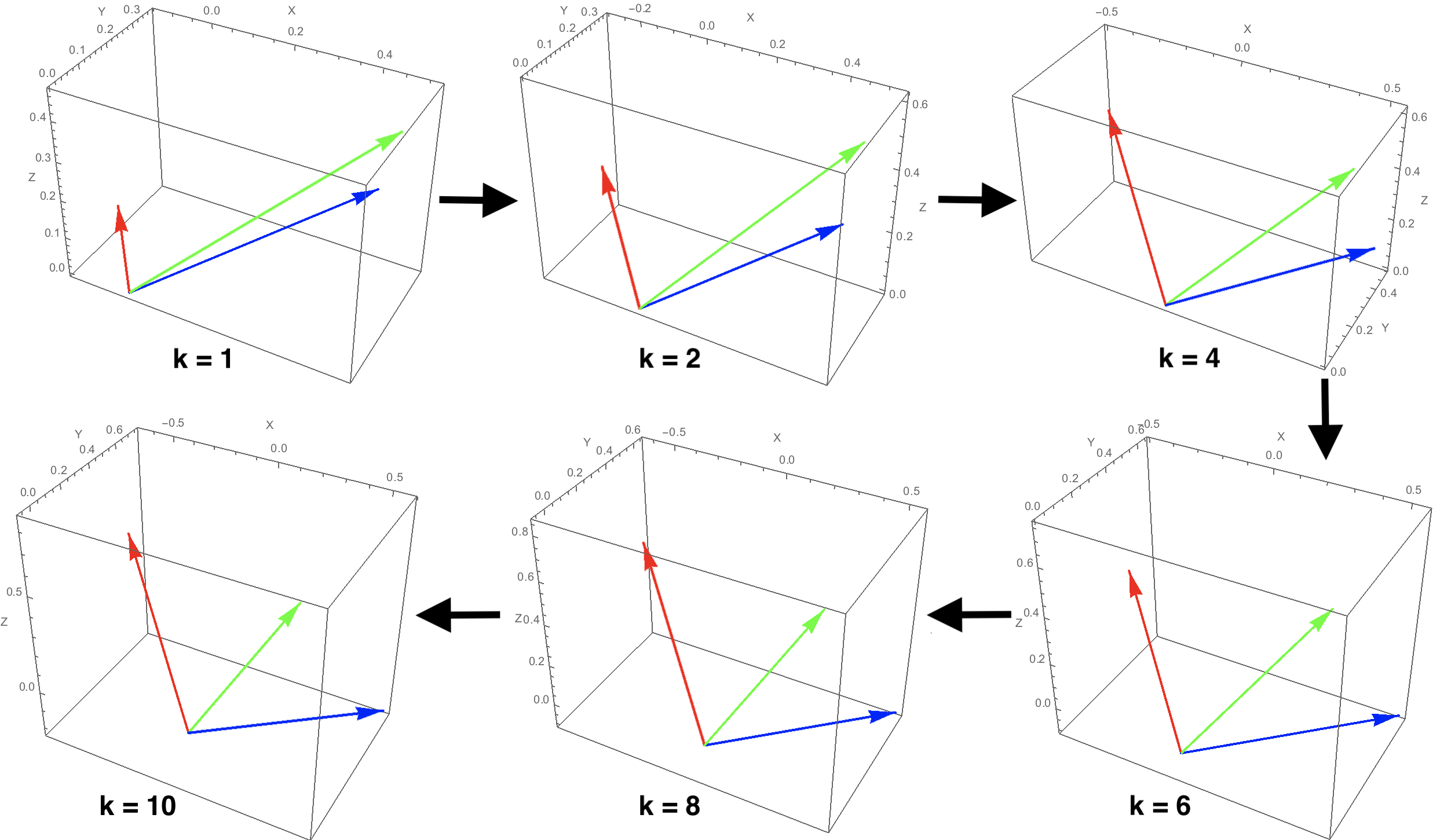}
\caption{The computed vectors of ${\bf{h}}_x, {\bf{h}}_y, {\bf{h}}_z$ through iteration (\ref{iteration}). The red, green and blue arrows stand for ${\bf{h}}_x, {\bf{h}}_y, {\bf{h}}_z$ respectively where $k$ is the iteration index.}
\label{fig:vectors}
\end{figure*}

\subsection{Further Discussion}
Letting
\begin{equation}
\begin{gathered}
  \kappa  =  {\left\| {{{\mathbf{h}}_x}} \right\|^2} + {\left\| {{{\mathbf{h}}_y}} \right\|^2} + {\left\| {{{\mathbf{h}}_z}} \right\|^2} \hfill \\
  \begin{array}{*{20}{c}}
  {}&{}&{} 
\end{array} + H_{x3}^2 + H_{y3}^2 + H_{z3}^2 \hfill \\
  \begin{gathered}
  {\cal{Y}} = {{\cal{H}}_1}\left( {{{\mathbf{h}}_x}} \right){{\cal{H}}_2}\left( {{{\mathbf{h}}_y}} \right) + {{\cal{H}}_2}\left( {{{\mathbf{h}}_y}} \right){{\cal{H}}_1}\left( {{{\mathbf{h}}_x}} \right) +  \hfill \\
  \begin{array}{*{20}{c}}
  {}&{}&{} 
\end{array}{{\cal{H}}_1}\left( {{{\mathbf{h}}_x}} \right){{\cal{H}}_3}\left( {{{\mathbf{h}}_z}} \right) + {{\cal{H}}_3}\left( {{{\mathbf{h}}_z}} \right){{\cal{H}}_1}\left( {{{\mathbf{h}}_x}} \right) +  \hfill \\
  \begin{array}{*{20}{c}}
  {}&{}&{} 
\end{array}{{\cal{H}}_2}\left( {{{\mathbf{h}}_y}} \right){{\cal{H}}_3}\left( {{{\mathbf{h}}_z}} \right) + {{\cal{H}}_3}\left( {{{\mathbf{h}}_z}} \right){{\cal{H}}_2}\left( {{{\mathbf{h}}_y}} \right) \hfill \\ 
\end{gathered}  \hfill \\ 
\end{gathered} 
\end{equation}
we have
\begin{equation}
{{\bf{W}}^2} = \kappa {\bf{I}} + {\cal{Y}}
\end{equation}
As $\kappa \bf{I}$ is commutative with any matrix, it is derived that
\begin{equation}
{{\bf{W}}^4} = {\left( {\kappa {\bf{I}} + {\cal{Y}}} \right)^2} = {\kappa ^2}{\bf{I}} + {\cal{Y}}(2\kappa {\bf{I}} + {\cal{Y}})
\end{equation}
and then
\begin{equation}
\begin{gathered}
  {{\mathbf{W}}^8} = {\left[ {{\kappa ^2}{\mathbf{I}} + {\cal{Y}}(2\kappa {\mathbf{I}} + {\cal{Y}})} \right]^2} \hfill \\
   = {\kappa ^4}{\mathbf{I}} + {\cal{Y}}(2\kappa {\mathbf{I}} + {\cal{Y}})\left[ {2{\kappa ^2}{\mathbf{I}} + {\cal{Y}}(2\kappa {\mathbf{I}} + {\cal{Y}})} \right] \hfill \\ 
\end{gathered} 
\end{equation}
Hence by letting
\begin{equation}
\begin{array}{l}
{{\cal{Y}}_1} = {\cal{Y}}\\
{{\cal{Y}}_2} = {{\cal{Y}}_1}(2\kappa {\bf{I}} + {{\cal{Y}}_1})\\
 \vdots \\
{{\cal{Y}}_j} = {{\cal{Y}}_{j - 1}}(2{\kappa ^{{2^{j - 2}}}}{\bf{I}} + {{\cal{Y}}_{j - 1}}),j = 2,3, \cdots 
\end{array}
\end{equation}
, we have
\begin{equation}
\begin{gathered}
  \mathop {\lim }\limits_{j \to  + \infty } {{\mathbf{W}}^{{2^j}}} 
   = \mathop {\lim }\limits_{j \to  + \infty } \left( {{\kappa ^{{2^{j - 1}}}}{\mathbf{I}} + {{\cal{Y}}_j}} \right) \hfill \\
   = \mathop {\lim }\limits_{j \to  + \infty } \left[ {{\kappa ^{{2^{j - 1}}}}{\mathbf{I}} + {{\cal{Y}}_{j - 1}}(2{\kappa ^{{2^{j - 2}}}} + {{\cal{Y}}_{j - 1}})} \right] \hfill \\ 
\end{gathered} 
\end{equation}
For ${\cal{Y}}_j$, its eigenvector $\bf{q}$ satisfies
\begin{equation}\label{B_eig_d}
\begin{gathered}
  {{\cal{Y}}_1}{\mathbf{q}} = ({{\mathbf{W}}^2} - \kappa {\mathbf{I}}){\mathbf{q}} = \left( {{\lambda_{\bf{W}} ^2} - \kappa } \right){\mathbf{q}} \hfill \\
  {{\cal{Y}}_2}{\mathbf{q}} = {{\cal{Y}}_1}(2\kappa {\mathbf{I}} + {{\cal{Y}}_1}){\mathbf{q}} \hfill \\
  \begin{array}{*{20}{c}}
  {}&{}&{} 
\end{array} = 2\kappa {{\cal{Y}}_1}{\mathbf{q}} + {\cal{Y}}_1^2{\mathbf{q}} = \left[ {2\kappa \left( {{\lambda_{\bf{W}} ^2} - \kappa } \right) + {{\left( {{\lambda_{\bf{W}} ^2} - \kappa } \right)}^2}} \right]{\mathbf{q}} \hfill \\
   \vdots  \hfill \\
  {{\cal{Y}}_j}{\mathbf{q}} = {{\cal{Y}}_{j - 1}}(2{\kappa ^{{2^{j - 2}}}}{\mathbf{I}} + {{\cal{Y}}_{j - 1}}){\mathbf{q}} \hfill \\
  \begin{array}{*{20}{c}}
  {}&{}&{} 
\end{array} = {\lambda _{{{\cal{Y}}_{j - 1}}}}\left( {2{\kappa ^{{2^{j - 2}}}} + {\lambda _{{{\cal{Y}}_{j - 1}}}}} \right){\mathbf{q}} = {\lambda _{{{\cal{Y}}_j}}}{\mathbf{q}} \hfill \\ 
\end{gathered} 
\end{equation}
where $\lambda_{{\cal{Y}}_j}$ is the eigenvalue of ${\cal{Y}}_j$. With the above evolutions, the eigenvalue of the power matrix ${{\bf{W}}^{2^j}}$ can be computed by
\begin{equation}\label{W_eig_1}
{\lambda _{{{\bf{W}}^{{2^j}}}}} = {\kappa ^{{2^{j - 1}}}} + {\lambda _{{{\cal{Y}}_{j - 1}}}}\left( {2{\kappa ^{{2^{j - 2}}}} + {\lambda _{{{\cal{Y}}_{j - 1}}}}} \right),j = 2,3, \cdots 
\end{equation}
From (\ref{B_eig_d}), one can find out that the eigenvalue of ${\cal{Y}}_j$ is in fact in a second-order recursive series, such that
\begin{equation}
\left\{ \begin{array}{l}
{\lambda _{{{\cal{Y}}_1}}} = {\lambda _{\cal{Y}}}\\
{\lambda _{{{\cal{Y}}_j}}} = {\lambda _{{{\cal{Y}}_{j - 1}}}}\left( {2{\kappa ^{{2^{j - 2}}}} + {\lambda _{{{\cal{Y}}_{j - 1}}}}} \right)
\end{array} \right.,j = 2,3, \cdots 
\end{equation}
The general solution to this series formula is given by
\begin{equation}
{\lambda_{{\cal{Y}}_j}} = {e^{{C_{init}}{2^{j - 1}}}} - {\kappa ^{{2^{j - 1}}}},j = 2,3, \cdots 
\end{equation}
where $C_{init}$ is the initial condition of ${\lambda_{{\cal{Y}}_j}}$ determined by the $\lambda_{{\cal{Y}}_1}$:
\begin{equation}
\begin{gathered}
  {\lambda _{{{\cal{Y}}_j}}}{|_{j = 1}} = {\lambda _{{{\cal{Y}}_1}}} = {e^{{C_{init}}}} - \kappa  \hfill \\
   \Rightarrow {C_{init}} = \ln \left( {{\lambda _{{{\cal{Y}}_1}}} + \kappa } \right) = \ln \lambda _{\mathbf{W}}^2 = 2\ln \lambda _{\mathbf{W}}^{} \hfill \\ 
\end{gathered} 
\end{equation}
Hence we arrive at
\begin{equation}\label{B_eig}
{\lambda _{{{\cal{Y}}_j}}} = \lambda _{\bf{W}}^{{2^j}} - {\kappa ^{{2^{j - 1}}}} ,j = 2,3, \cdots 
\end{equation}
From the solving process we can see that the high-order eigenvalues are in the form of exponential function. This indirectly indicates that the system (\ref{iteration}) has exponential-like results. However, for (\ref{iteration}), the steady-state solution of the differential equations can hardly be computed. In fact the system (\ref{iteration}) indicates a solution set of the two independent systems
\begin{equation}
\begin{array}{*{20}{c}}
  \begin{gathered}
  {{\mathbf{h}}_{x,k}} = {\rho _{k - 1}}{{\mathbf{h}}_{x,k - 1}} \hfill \\
  {{\mathbf{h}}_{y,k}} = {\rho _{k - 1}}{{\mathbf{h}}_{y,k - 1}} \hfill \\
  {{\mathbf{h}}_{z,k}} = {\rho _{k - 1}}{{\mathbf{h}}_{z,k - 1}} \hfill \\ 
\end{gathered}  
\end{array},\begin{array}{*{20}{c}}
  \begin{gathered}
  {{\mathbf{h}}_{x,k}} = {\rho _{k - 1}}\left( {{{\mathbf{h}}_{y,k - 1}} \times {{\mathbf{h}}_{z,k - 1}}} \right) \hfill \\
  {{\mathbf{h}}_{y,k}} = {\rho _{k - 1}}\left( {{{\mathbf{h}}_{z,k - 1}} \times {{\mathbf{h}}_{x,k - 1}}} \right) \hfill \\
  {{\mathbf{h}}_{z,k}} = {\rho _{k - 1}}\left( {{{\mathbf{h}}_{x,k - 1}} \times {{\mathbf{h}}_{y,k - 1}}} \right) \hfill \\ 
\end{gathered}  
\end{array}
\end{equation}
Both of the solutions are hard to be figured out since $\rho_{k-1}$ here is nonlinear. Besides, the closed-form solution to this system may be very complicated, which loses its exact meaning for fast computation. But the analytical solution to (\ref{iteration}) is still worthy of research in the future.

\section{Experimental Results}
In this section, several simulations are carried out. The proposed method here is named after the Fast Analytical 3D Registration (FA3R). The comparisons on the accuracy, robustness and computation speed of the proposed FA3R with representative ones are presented. The general point cloud correspondences are simulated via the following model
\begin{equation}
{{\mathbf{b}}_i} = {{\mathbf{C}}_{true}}{{\mathbf{r}}_i} + {{\mathbf{T}}_{true}} + {\varepsilon _i}
\end{equation}
in which ${{\mathbf{C}}_{true}}$ and ${{\mathbf{T}}_{true}}$ are the true transformation parameters whose effects are corrupted by the noise item ${\varepsilon _i}$. Here for the sake of generality and convenience, the noise is refined to follow the normal distribution such that ${\varepsilon _i} \sim {\cal{N}}({\mathbf{0}},{{\bf{\Sigma}} _{{\varepsilon _i}}})$ where $\bf{\Sigma}$ denotes the covariance and here is assumed to be diagonal. The signal-to-noise ratio (SNR) is defined to be
\begin{equation}
{\rm{SNR}} = \frac{{\left\| {{\mathbf{\bar r}}} \right\|}}{{\left\| {{\mathbf{\bar b}} - {{\mathbf{T}}_{true}}} \right\|}}
\end{equation}
Note that in the following comparisons the weights are equalized for us to pay special attention to the internal of registration algorithms.

\begin{figure*}[ht]
\centering
\includegraphics[width=0.95\textwidth]{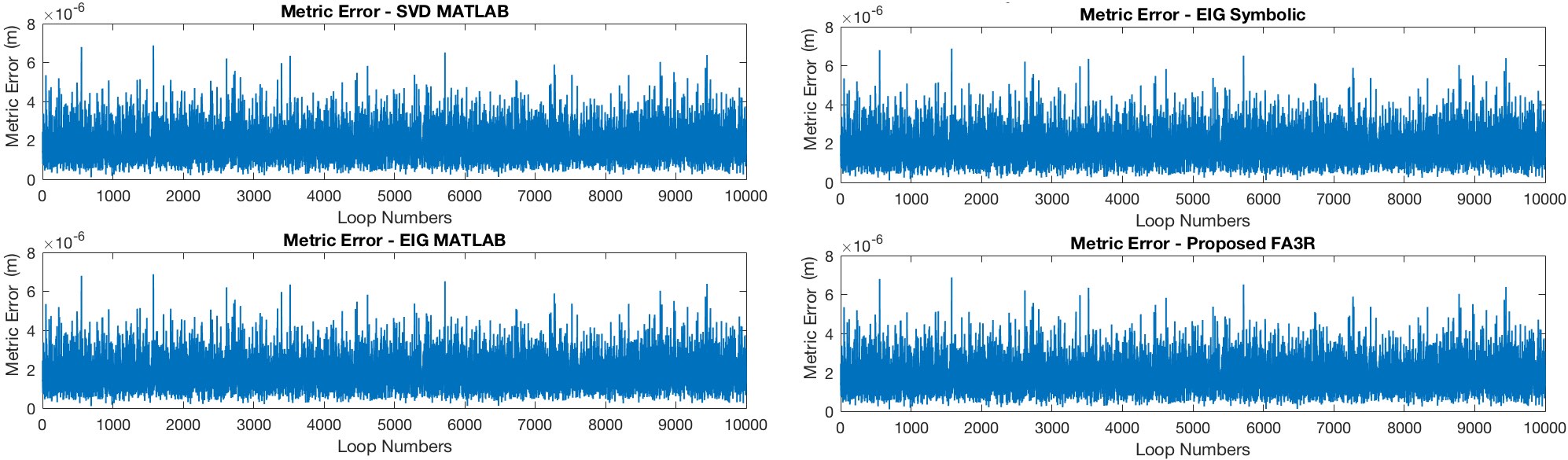}
\caption{The metric errors of various methods implemented using MATLAB, where the SNR $=1000$.}
\label{fig:metric_error}
\end{figure*}

\setcounter{figure}{3}
\begin{figure*}[hb]
\begin{center}
\begin{minipage}[c]{0.49\textwidth}
\centering
\includegraphics[width=1.0\textwidth]{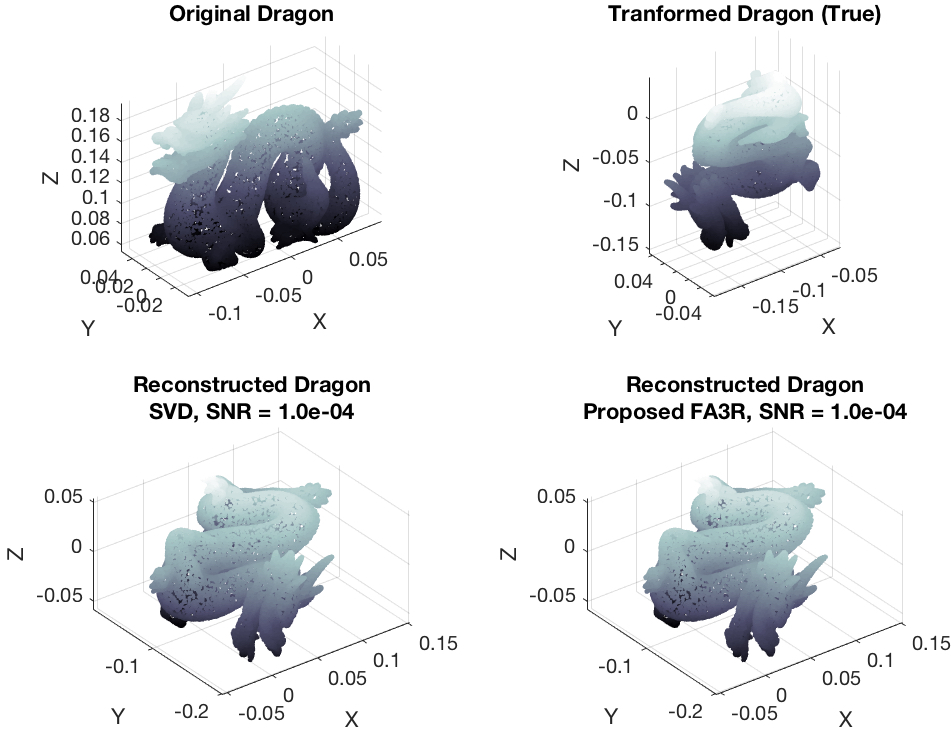}
\centering
\caption{3D reconstruction of a dragon from transformed point correspondences, where the SNR $=0.0001$.}
\label{fig:dragon_SNR_00001}
\end{minipage}%
\begin{minipage}[c]{0.48\textwidth}
\centering
\includegraphics[width=1.0\textwidth]{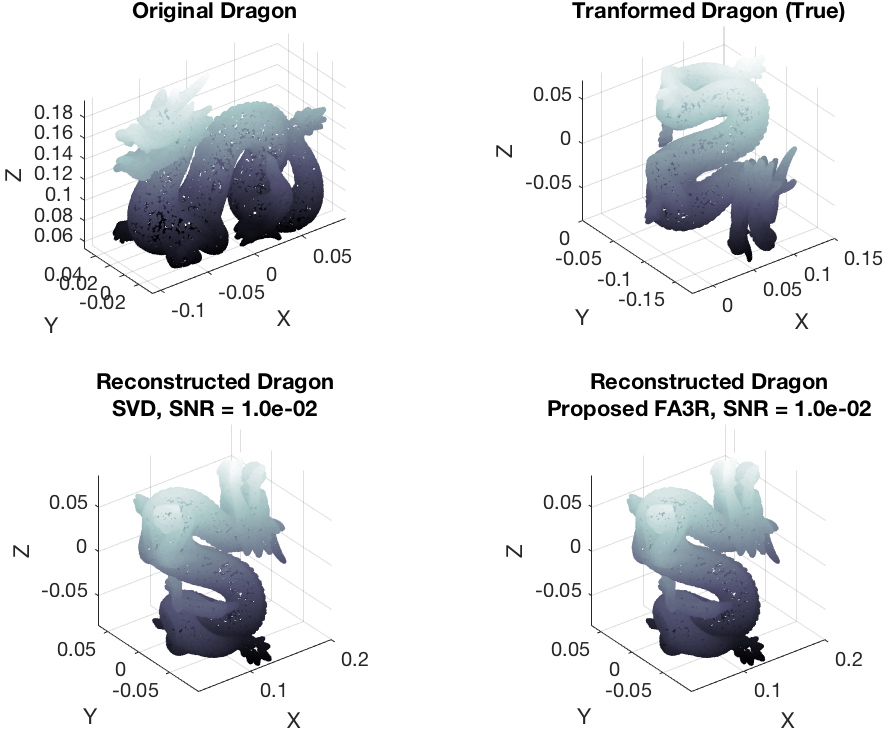}
\caption{3D reconstruction of a dragon from transformed point correspondences, where the SNR $=0.01$.}
\label{fig:dragon_SNR_001}
\end{minipage}
\end{center}
\end{figure*} 

\subsection{Accuracy, Convergence and Robustness}
In this sub-section, the point clouds are transformed by the following rotation and translation
\begin{equation}
\begin{gathered}
  {\mathbf{C}} = \left( {\begin{array}{*{20}{c}}
  {0.28660470}&{0.13449407}&	{0.94856158}\\	
{0.88489615}&{-0.41661863}&{-0.20829714}\\	
{0.36717370}&{0.89907744}&{-0.23841815} 
\end{array}} \right) \hfill \\
  {\mathbf{T}} = {\left( {6.2, - 8.7,4.3} \right)^T} \hfill \\ 
\end{gathered} 
\end{equation}
We generate 100 point vector pairs where the noise density is chosen to ensure the signal-noise-ratio (SNR) $=0.1$ so that the modeled point cloud contains heavily corrupted noises. The SVD by Umeyama and EIG from \cite{Umeyama1991} and \cite{Besl1992} are used for comparison respectively. The rotation matrix and metric error are mainly compared. The $\bf{D}$ matrix is obtained by
\begin{equation}
{\bf{D}} = \left( {\begin{array}{*{20}{c}}
{-0.1493707}&{0.33704186}&{-0.26092604}\\
{0.15536306}&{-0.15098108}&{0.87009800}\\
{0.72649274}&{-0.26632189}&{-0.91058475}
\end{array}} \right)
\end{equation}
Using SVD, we obtain the following rotation
\begin{equation}
{\bf{C}}_{{\rm{SVD}}} = \left( {\begin{array}{*{20}{c}}
{-0.0741384}&{0.99184073}&{0.10370840}\\
{0.15408691}&{-0.09135253}&{0.98382515}\\
{0.98527189}&{0.08891942}&{-0.14605694}
\end{array}} \right)	
\end{equation}
The eigen-decomposition result is shown as follows
\begin{equation}
{\bf{C}}_{{\rm{EIG}}} = \left( {\begin{array}{*{20}{c}}
{-0.07413848}&{0.99184073}&{0.10370840}\\
{0.15408691}&{-0.09135253}&{0.98382515}\\
{0.98527189}&{0.08891942}&{-0.14605694}
\end{array}} \right)	
\end{equation}
The proposed FA3R here is conducted with 10 iterations without halt determined by relative accuracy. With ${{\mathbf{h}}_x},{{\mathbf{h}}_y},{{\mathbf{h}}_z}$ extracted from $\bf{D}$, the iteration evolutes with general trajectory of
\begin{subequations}\label{evo}
\begin{equation}
\begin{gathered}
  {{\mathbf{h}}_{x,2}} = ({0.14284188,0.31721893,0.56072631)} \hfill \\
  {{\mathbf{h}}_{y,2}} = ({0.66185732,0.11521721, - 0.09376886)} \hfill \\
  {{\mathbf{h}}_{z,2}} = ({- 0.10369554,0.61094186, - 0.52616537)} \hfill \\
   \vdots  \hfill \\
  {{\mathbf{h}}_{x,5}} = ({0.10622539,0.58055939,0.80725691)} \hfill \\
  {{\mathbf{h}}_{y,5}} = ({0.98078912,0.07239918, - 0.18112799)} \hfill \\
  {{\mathbf{h}}_{z,5}} = ({- 0.16360071,0.81099106, - 0.56171856)} \hfill \\
   \vdots  \hfill \\
    \end{gathered}
\end{equation}
\begin{equation}
\begin{gathered}
  {{\mathbf{h}}_{x,10}} = ({0.10622550,0.58056084,0.80725785)} \hfill \\
  {{\mathbf{h}}_{y,10}} = ({0.98079096,0.07239924, - 0.18112822)} \hfill \\
  {{\mathbf{h}}_{z,10}} = ({ - 0.16360081,0.81099164, - 0.56171818)} \hfill \\ 
\end{gathered} 
\end{equation}
\end{subequations}

\noindent The FA3R finally gives the rotation of
\begin{equation}
{\bf{C}}_{{\rm{FA3R}}} = \left( {\begin{array}{*{20}{c}}
{-0.07413848}&{0.99184073}&{0.10370840}\\
{0.15408691}&{-0.09135253}&{0.98382515}\\
{0.98527189}&{0.08891942}&{-0.14605694}
\end{array}} \right)	
\end{equation}
which is exactly the same with ${\bf{C}}_{{\rm{SVD}}}$ and ${\bf{C}}_{{\rm{EIG}}}$. In this case, the metric error values for the SVD, EIG and proposed FA3R are
\begin{equation}
\begin{gathered}
  {L_{{\rm{SVD}}}} = 10.31257252111700539743 \hfill \\
  {L_{{\rm{EIG}}}} = 10.31257252111700550845 \hfill \\
  {L_{{\rm{FA3R}}}} = 10.31257252111700534192 \hfill \\ 
\end{gathered} 
\end{equation}
\begin{figure*} [ht]
\begin{center}
\begin{minipage}[c]{0.48\textwidth}
\centering
\includegraphics[width=1.0\textwidth]{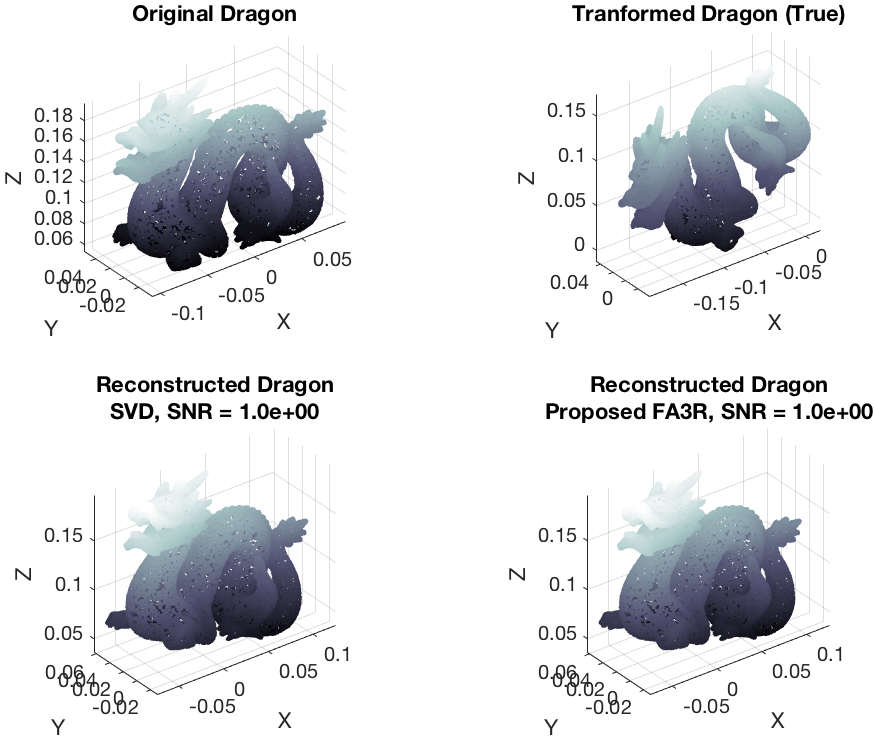}
\centering
\caption{3D reconstruction of a dragon from transformed point correspondences, where the SNR $=1$.}
\label{fig:dragon_SNR_1}
\end{minipage}%
\begin{minipage}[c]{0.48\textwidth}
\centering
\includegraphics[width=1.0\textwidth]{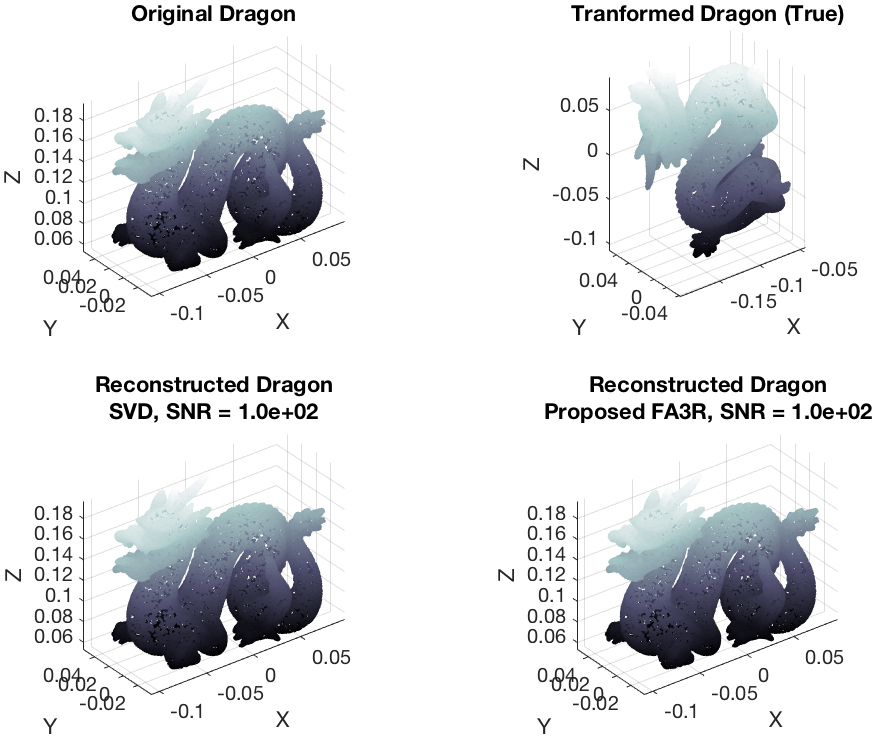}
\caption{3D reconstruction of a dragon from transformed point correspondences, where the SNR $=100$.}
\label{fig:dragon_SNR_100}
\end{minipage}
\end{center}
\end{figure*} 

\begin{table*}[hb]
\centering
\caption{Studied Cases for Robustness Verification}\label{tab:cases}
\resizebox{1.0\textwidth}{!}{
\begin{tabular}{cccccc}
\toprule
{Case}&{True Euler Angles $\varphi ,\vartheta ,\psi $}&{True Translation $\bf{T}$}&{Noise Covariance ${{\bf{\Sigma}} _{{\eta _i}}}$}&{Vector Number $n$}&{$rank({\bf{D}})$}\\
\midrule
{1}&{$\left( { - \frac{\pi }{6},\frac{{4\pi }}{{11}}, - \frac{{5\pi }}{7}} \right)=(-0.52359878, 1.1423973, -2.2439948)$}&{${\left( {100, - 50,80} \right)^T}$}&{$diag\left( {0.0,0.0,0.0} \right)$}&{100}&{3}\\
{2}&{$\left( { - \frac{\pi }{6},\frac{{4\pi }}{{11}}, - \frac{{5\pi }}{7}} \right)=(-0.52359878, 1.1423973, -2.2439948)$}&{${\left( {100, - 50,80} \right)^T}$}&{$diag\left( {0.0,0.0,0.0} \right)$}&{100}&{2}\\
{3}&{$\left( { - \frac{\pi }{6},\frac{{4\pi }}{{11}}, - \frac{{5\pi }}{7}} \right)=(-0.52359878, 1.1423973, -2.2439948)$}&{${\left( {100, - 50,80} \right)^T}$}&{$diag\left( {0.0,0.0,0.0} \right)$}&{100}&{1}\\
{4}&{$\left( { \frac{4 \pi }{7},\frac{{\pi }}{{2}}, - \frac{{9\pi }}{20}} \right)=(1.7951958, 1.5707963, -1.4137167)$}&{${\left( {-60, 70, 40} \right)^T}$}&{$diag\left( {10,10,10} \right)$}&{100}&{3}\\
{5}&{$\left( { \frac{4 \pi }{7},\frac{{\pi }}{{2}}, - \frac{{9\pi }}{20}} \right)=(1.7951958, 1.5707963, -1.4137167)$}&{${\left( {-60, 70, 40} \right)^T}$}&{$diag\left( {10,10,10} \right)$}&{1000}&{3}\\
{6}&{$\left( { \frac{4 \pi }{7},\frac{{\pi }}{{2}}, - \frac{{9\pi }}{20}} \right)=(1.7951958, 1.5707963, -1.4137167)$}&{${\left( {-60, 70, 40} \right)^T}$}&{$diag\left( {10,10,10} \right)$}&{10000}&{3}\\
{7}&{$\left( { \frac{5 \pi }{9}, - \frac{{7\pi }}{{10}}, \frac{{4\pi }}{13}} \right)=(-1.3962634, -0.9424778, -2.1749488)$}&{${\left( {80, -20, -160} \right)^T}$}&{$diag\left( {0.1,10,1000} \right)$}&{1000}&{3}\\
{8}&{$\left( { \frac{5 \pi }{9}, - \frac{{7\pi }}{{10}}, \frac{{4\pi }}{13}} \right)=(-1.3962634, -0.9424778, -2.1749488)$}&{${\left( {80, -20, -160} \right)^T}$}&{$diag\left( {1000,10,0.1} \right)$}&{1000}&{3}\\
{9}&{$\left( { \frac{5 \pi }{9}, - \frac{{7\pi }}{{10}}, \frac{{4\pi }}{13}} \right)=(-1.3962634, -0.9424778, -2.1749488)$}&{${\left( {80, -20, -160} \right)^T}$}&{$diag\left( {0.1,0,1,0.1} \right)$}&{1000}&{3}\\
\bottomrule
\\
\bottomrule
\end{tabular}}
\end{table*}
\noindent which indicate that FA3R here is even better with a very slightly smaller value. We can see from (\ref{evo}) that the iterated results at indices $k=5$ and $k=10$ have quite small differences. In fact, when the relative accuracy $\epsilon$ is set to $1\times10^{-14}$, the iteration stops at the index of $k=6$. Generally, 
\setcounter{figure}{2}
\begin{figure}[H]
\centering
\includegraphics[width=0.5\textwidth]{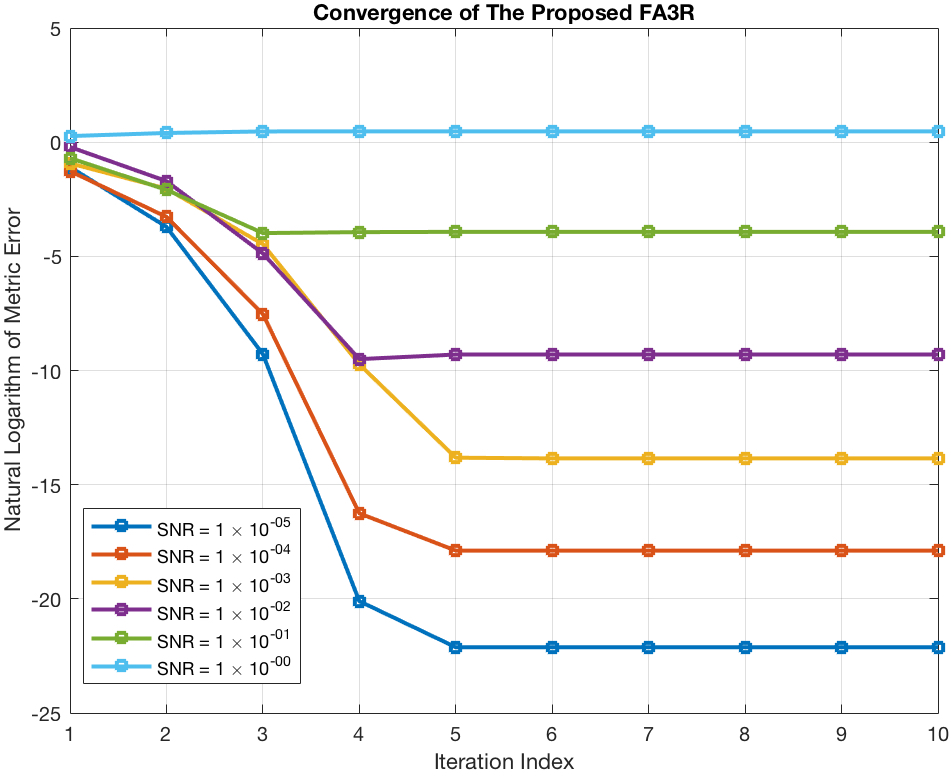}
\caption{The convergence performance of the proposed FA3R where the $y$-axis is the natural logarithm of the metric error.}
\label{fig:convergence}
\end{figure}
\noindent the FA3R, SVD and EIG have the same accuracy and robustness. This is because they are based on the same least-square framework in (\ref{opt}). A point cloud case with SNR $= 1000$ is simulated which generates the metric error evaluation in Fig. \ref{fig:metric_error}. In this simulation, the stopping threshold is set to $1\times10^{-12}$. The metric errors of SVD, EIG and FA3R show very similar results which coincide with previous analysis. Another case is then generated to demonstrate the calculation process. The vectors ${\bf{h}}_x, {\bf{h}}_y, {\bf{h}}_z$ in iteration are shown in Fig. \ref{fig:vectors} in which we can see that with evolution the vectors gradually become orthonormal to each other in the final steps. This verifies the findings in Section \ref{sec:evo} that the final iterated ${\bf{h}}_x, {\bf{h}}_y, {\bf{h}}_z$ constitute the orthonormal bases of the optimal DCM. The convergence of the proposed FA3R is studied via 6 simulations where the SNRs are differently given. The relationships between the iteration number and convergency are shown in Fig. \ref{fig:convergence}. When processing the point cloud with low magnitude of noises, the algorithm will converge very closely to 0 within 5 iterations. For more noisy cases, the convergence is worse for a little bit, which require almost $6 \sim 8$ iterations to converge.\\
\indent With the dragon example presented by Computer Graphics Laboratory, Stanford University \cite{Curless1996A}, we are able to do 3D reconstruction of a given dragon point cloud measurement. The rotation and translation are estimated by SVD and proposed FA3R that map the original point cloud to transformed ones (see Fig. \ref{fig:dragon_SNR_00001}, \ref{fig:dragon_SNR_001}, \ref{fig:dragon_SNR_1} and \ref{fig:dragon_SNR_100}). In the four scenarios, the transformed dragon is rotated by random Euler angles and translated randomly. The SNRs are set to $0.0001$, $0.01$, $1$ and $100$ respectively generating different magnitudes of stochastic noises. The tested reconstruction results show that the proposed FA3R reaches the same accuracy with that of SVD. The studied case is polluted in part by large-magnitude noises. In real engineering practice, when conduct 3D registration based on 3D LiDAR measurements, the SNR is very high and reach up to $100 \sim 10000$. In such condition, only $2 \sim 4$ iterations are required for FA3R to compute adequately accurate values for double-precision processing.\\

\begin{table*}[ht]
\centering
\caption{Estimated Euler Angles $\varphi ,\vartheta ,\psi $}\label{tab:euler}
\resizebox{0.65\textwidth}{!}{
\begin{tabular}{cccccc}
\toprule
{Case}&{SVD}&{EIG}&{Proposed FA3R}\\
\midrule
{1}&{$(-0.5235, 1.1424, -2.2439)$}&{$(-0.5235, 1.1424, -2.2439)$}&{$(-0.5235, 1.1424, -2.2439)$}\\
{2}&{$(-2.8874, 0.6156, -2.3558)$}&{$(1.3088, 0.6156, -2.3558)$}&{$(1.3088, 0.6156, -2.3558)$}\\
{3}&{$(-0.04696, -0.04481, -0.04696)$}&{$(-2.0344, 0.7297, -2.0344)$}&{$(-2.0344, 0.7297, -2.0344)$}\\
{4}&{$(0.3614, 1.258, -0.5719)$}&{$(0.3614, 1.258, -0.5719)$}&{$(0.3614, 1.258, -0.5719)$}\\
{5}&{$(0.6665, 1.4052, -0.4474)$}&{$(0.6665, 1.4052, -0.4474)$}&{$(0.6665, 1.4052, -0.4474)$}\\
{6}&{$(-0.02599, 1.4942, 0.4477)$}&{$(-0.02599, 1.4942, 0.4477)$}&{$(-0.02599, 1.4942, 0.4477)$}\\
{7}&{$(2.7465, 0.5139, 2.7899)$}&{$(2.7465, 0.5139, 2.7899)$}&{$(2.7465, 0.5139, 2.7899)$}\\
{8}&{$(0.2577, -0.3486, 0.2181)$}&{$(0.2577, -0.3486, 0.2181)$}&{$(0.2577, -0.3486, 0.2181)$}\\
{9}&{$(-1.4018, -0.9443, -2.1804)$}&{$(-1.4018, -0.9443, -2.1804)$}&{$(-1.4018, -0.9443, -2.1804)$}\\
\bottomrule
\\
\bottomrule
\end{tabular}}

\centering
\caption{Estimated Translation $\bf{T}$}\label{tab:trans}
\resizebox{0.65\textwidth}{!}{
\begin{tabular}{cccccc}
\toprule
{Case}&{SVD}&{EIG}&{Proposed FA3R}\\
\midrule
{1}&{$(100.015, -50.0834, 79.9858)^T$}&{$(100.015, -50.0834, 79.9858)^T$}&{$(100.015, -50.0834, 79.9858)^T$}\\
{2}&{$(99.8532, -49.9929, -49.9929)^T$}&{$(99.8532, -49.9929, -49.9929)^T$}&{$(99.8532, -49.9929, -49.9929)^T$}\\
{3}&{$(100.0, 100.0, 100.0)^T$}&{$(100.0, 100.0, 100.0)^T$}&{$(100.0, 100.0, 100.0)^T$}\\
{4}&{$(-59.3406, 69.5444, 39.2757)^T$}&{$(-59.3406, 69.5444, 39.2757)^T$}&{$(-59.3406, 69.5444, 39.2757)^T$}\\
{5}&{$(-59.8461, 69.6513, 40.1395)^T$}&{$(-59.8461, 69.6513, 40.1395)^T$}&{$(-59.8461, 69.6513, 40.1395)^T$}\\
{6}&{$(-59.8461, 69.6513, 40.1395)^T$}&{$(-59.8461, 69.6513, 40.1395)^T$}&{$(-59.8461, 69.6513, 40.1395)^T$}\\
{7}&{$(79.9458, -19.9293, 141.043)^T$}&{$(79.9458, -19.9293, 141.043)^T$}&{$(79.9458, -19.9293, 141.043)^T$}\\
{8}&{$(91.9475, -20.049, 160.038)^T$}&{$(91.9475, -20.049, 160.038)^T$}&{$(91.9475, -20.049, 160.038)^T$}\\
{9}&{$(79.9251, -20.0097, 159.997)^T$}&{$(79.9251, -20.0097, 159.997)^T$}&{$(79.9251, -20.0097, 159.997)^T$}\\
\bottomrule
\\
\bottomrule
\end{tabular}}
\end{table*}

\begin{table*}[hb]
\centering
\caption{MATLAB execution time comparisons (SNR = 10)}\label{tab:MATLAB_time}
\resizebox{0.65\textwidth}{!}{
\begin{tabular}{cccccc}
\toprule
{Algorithms}&{Test 1}&{Test 2}&{Test 3}&{Test 4}&{Test 5}\\
\midrule
{SVD}&{0.319724423 s}&{0.335187723 s}&{0.332208797 s}&{0.311607620 s}&{0.342420298 s}\\
{EIG}&{0.351826581 s}&{0.35335691 s}&{0.354455200 s}&{0.329191103 s}&{0.328164580 s}\\
{Proposed FA3R}&{0.152435233 s}&{0.14824660 s}&{0.152318794 s}&{0.143857752 s}&{0.138070595 s}\\
\bottomrule
\end{tabular}}
\end{table*}

\indent As far as the robustness is concerned, the SVD in fact has analytical version that is recently proposed by us \cite{Liu2018}. However, such method has been proved to be unstable on robustness in real applications \cite{Wu2018ase}. However, the FA3R, EIG, and recent symbolic EIG \cite{Wu2018ase} and improved one \cite{Wu2018fast} all employ the same decomposition framework which is regarded and verified to be as robust as SVD. Here we use the previous dataset designed especially for robustness verification (\cite{Wu2018fast}, see Table \ref{tab:cases}). In this table the reference Euler angles $\varphi ,\vartheta ,\psi $ i.e. roll, pitch and yaw along with true translation vectors and other settings are given. The calculated results are shown in Table \ref{tab:euler} and \ref{tab:trans}. The results reveal that the proposed method owns the same accuracy and robustness regarding different configurations. The reason is obvious because the proposed FA3R has the same least-square model with conventional ones e.g. SVD, EIG and their variants. Therefore FA3R has the same limit of precision with these ones in computation.

\setcounter{figure}{7}
\begin{figure*}[ht]
\centering
\includegraphics[width=1.0\textwidth]{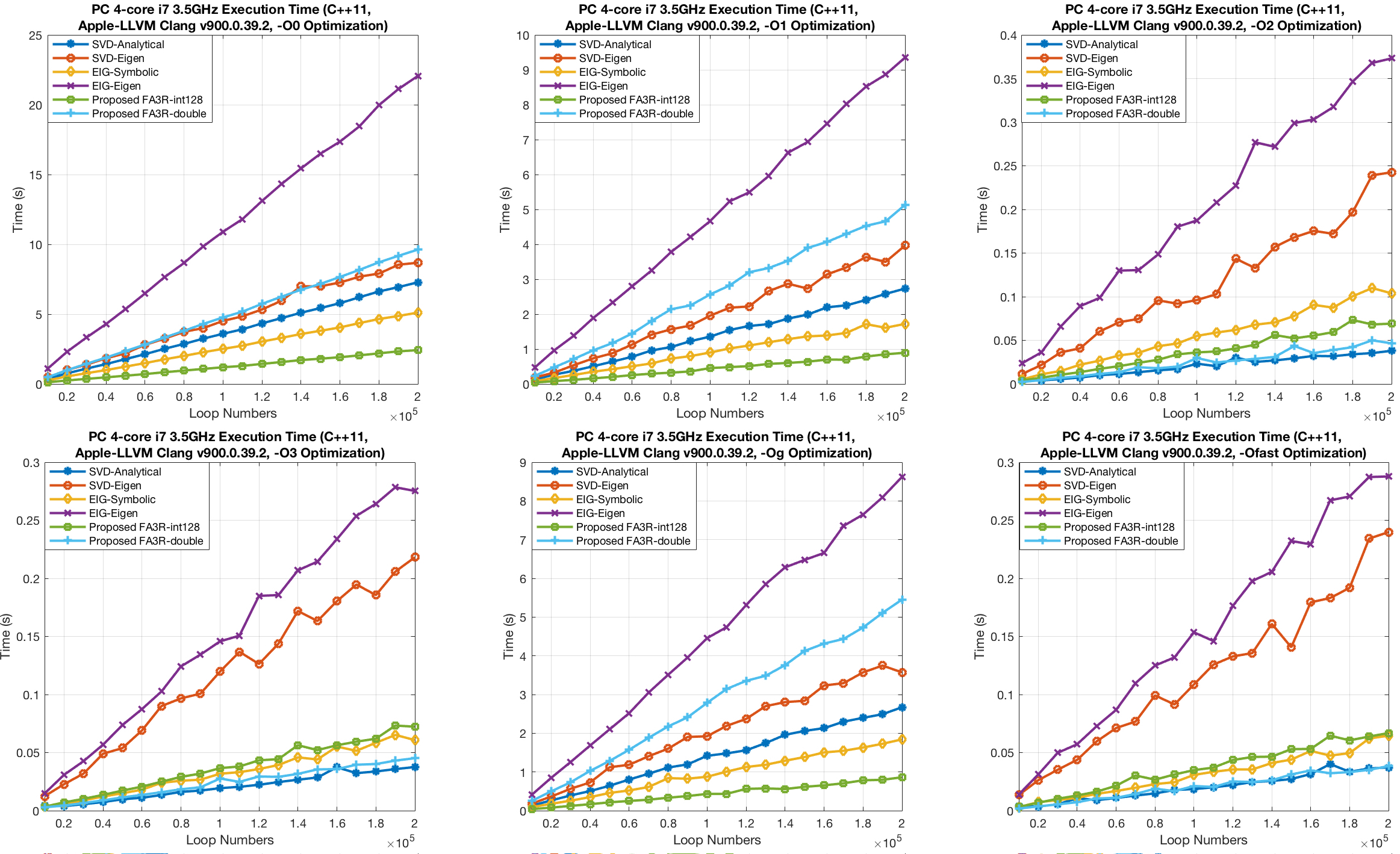}
\caption{Execution time complexity by C++ implementation on the PC with various compiling optimization parameters and increasing iteration numbers.}
\label{fig:time_PC}
\end{figure*}

\subsection{Execution Time Complexity}
This paper has a main claim on the time consumption advance in real computation. Here, the algorithms are implemented first using MATLAB r2016 and then they are rewritten in C++ programming language where the numerical procedures like SVD, determinant, eigenvalue decomposition are offered by the Eigen mathematical library which is one of the most popular numerical tools. The algorithms of SVD, eigenvalue decomposition, FA3R and internal computations are simplified to their own best ensuring fairness. A MacBook Pro 2017 Pro computer with 4-core i7 CPU of 3.5GHz clock speed is used for algorithm validation. Each algorithm is executed for 10000 times and the sum of the time is recorded by the internal timer. For MATLAB implementation, the SVD is implemented using MATLAB internal function $svd$ and the EIG employs the symbolic solution given in \cite{Wu2018ase} for maximum speed. We only logged the statistics of the registration computation as the data preparation parts are the same for all the algorithms. \\
\indent We can see from the Table \ref{tab:MATLAB_time} that the proposed FA3R method consumes nearly less than 50\% of SVD or eigenvalue decomposition. What needs to be pointed out is that the SVD in MATLAB has been extensively optimized and executes with very fast speed. The symbolic method in \cite{Wu2018} is the known fastest solving method for such eigenvalue problem. This indicatess even faster computation speed using compiling-run coding. Next, we rewrite the codes of SVD \cite{Umeyama1991}, fast analytical SVD \cite{Liu2018}, EIG \cite{Besl1992}, improved symbolic EIG \cite{Wu2018fast} and our proposed FA3R for PC and embedded platforms using C++. We first generate a case where the SNR is 10 and it contains 10000 matched point correspondences. The compilation of the program using C/C++ is very sensitive to the optimization levels offered by the compiler. In this paper, all the programs are evaluated using the most famous GCC compiler and its variant Clang. The matrix operations e.g. multiplication, SVD, EIG are implemented using the famous Eigen library with C++11 standard. The proposed FA3R is implemented using both the integer and double version. For PC i.e. the Apple MacBook Pro laptop, the Clang compiler is set as default which owns 6 basic optimization options. For most reliability-oriented works, the programmers tend not to use high-level optimizations since they usually induce instability and crash according to details like word alignment, unused symbols, improper memory usage, optimized jumps and etc. O0, O1, O2, O3 are 4 basic levels while Og and Ofast are optimized for faster compiling and faster runtime computation time respectively. All these options are enabled respectively to produce in-run time consumption results which are presented in Fig. \ref{fig:time_PC}.

\begin{figure*}[ht]
\centering
\includegraphics[width=1.0\textwidth]{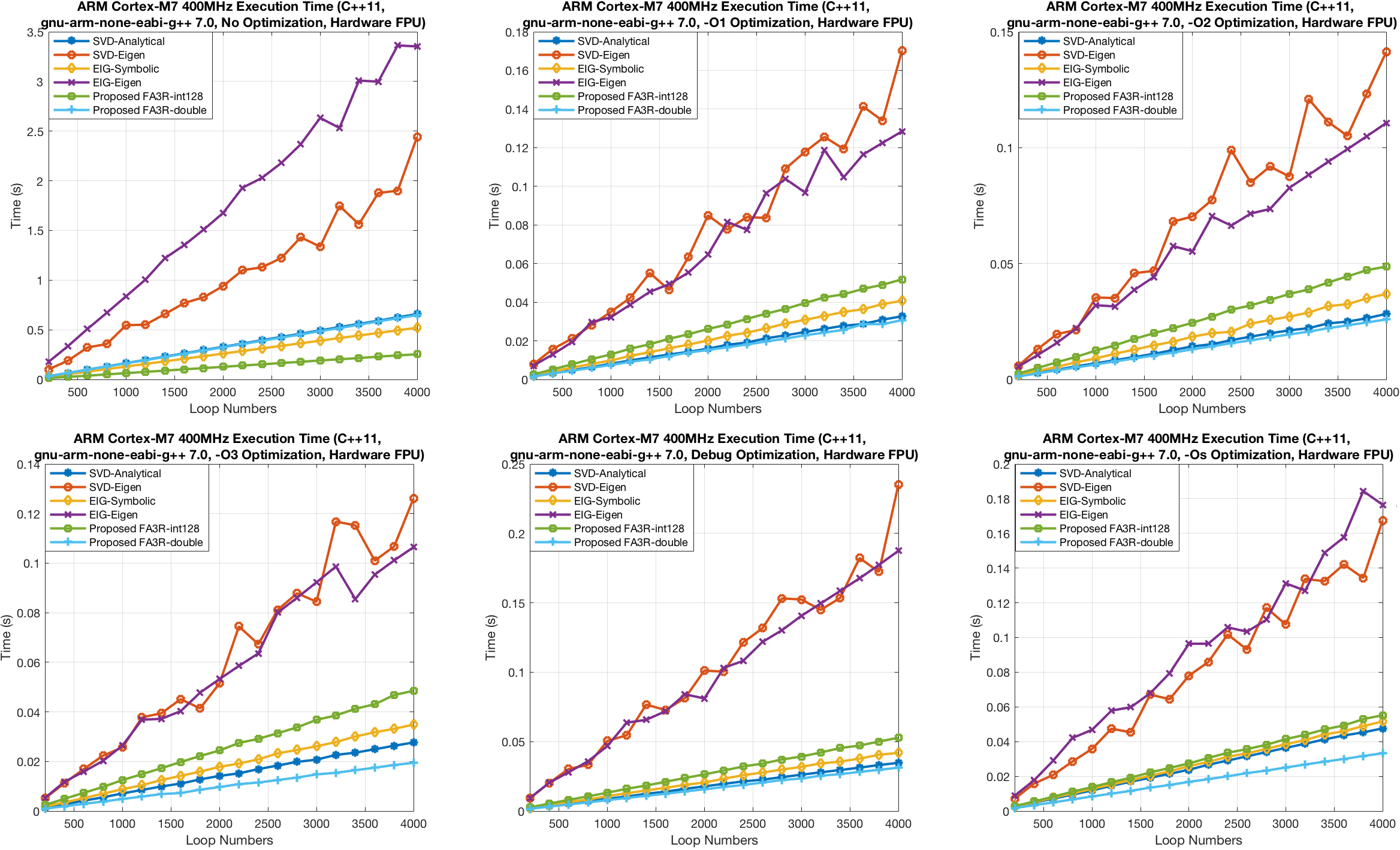}
\caption{Execution time complexity by C++ implementation on the STM32H743VIT6 (With FPU) containing various compiling optimization parameters and increasing iteration numbers.}
\label{fig:time_ARM_Hard_FPU}
\end{figure*}

\indent All the algorithms own the time complexity of $O(n)$ that consumes linearly as the loop number increases. For low-level optimizations like O0, O1 and Og, the algorithmic behaviors as quite stable. But this has been broken since the utilization of high-level optimization options. Especially, for Eigen-based algorithms, their computation time differs at relative large scale. One can also see from Fig. \ref{fig:time_PC} that sometimes the analytical SVD will be faster than FA3R but it has already been proved that this method has bad performance on robustness and accuracy \cite{Wu2018ase}. So in general the proposed FA3R owns the best balance among all the algorithms.\\
\indent Then the experiment is transferred to the embedded platform where an STM32H743VIT6 ARM Cortex-M7 micro controller with 1MB RAM and core clock speed of 400MHz is mounted. According to limited RAM space and computation speed, there are only 200 points for each validation and each algorithm is executed for 1000 times for average time evaluation. The codes are compiled using the GCC gnu-arm-none-eabi-g++ 7.0 with C++11 and Eigen library. The time consumption is calculated by means of the internal high-resolution 32-bit hardware timer. The adopted STM32H743VIT6 has a hardware floating-number processing unit (FPU). First the FPU is enabled for real engineering performance validation. The tested results are summarized in Fig. \ref{fig:time_ARM_Hard_FPU}. The performances are quite similar with that on PC but are more consistent because the program on PC is usually scheduled using multi-core CPU. The micro controller has only one core causing the very precise timing. When the FPU is used, the floating-number processing is conducted at fantastic speed. In such occasion the fixed-point version of FA3R would not always be better than double version. But on the platforms like FPGA and GPU there no way that the floating numbers can be directly accessed, according to the current technology. In such case the fixed-point version of FA3R would beat most of the algorithms with dependencies on numerical operations. The FPGA's time evaluation can hardly be processed via real-world hardware. Simulation results of FPGA are not usually reliable according to some impossible synthesis of circuit \cite{Quang2014,Guo2013}. At this point, we simulate the fixed-point behaviors via the micro controller without FPU. In this mode, the micro controller can only computer the floating numbers via software approximations which is regarded as very low-speed. The results are depicted in Fig. \ref{fig:time_ARM_No_FPU} where one can see that in every case the fixed-point version has much better performance than others. The average advance of the double version of FA3R in presented demonstrations is $67.74\%$ while for the fixed-point version it reaches $76.72\%$ at least and $86.01\%$ at most. The SVD and proposed fixed-point version FA3R are translated into the Verilog hardware description language. While the SVD consumes 15 times gates that of the FA3R. This indicates that the FA3R can not only lower the execution time load but can significantly decrease the power consumption in engineering implementations.

\setcounter{figure}{11}
\begin{figure*}[ht]
\centering
\includegraphics[width=1.0\textwidth]{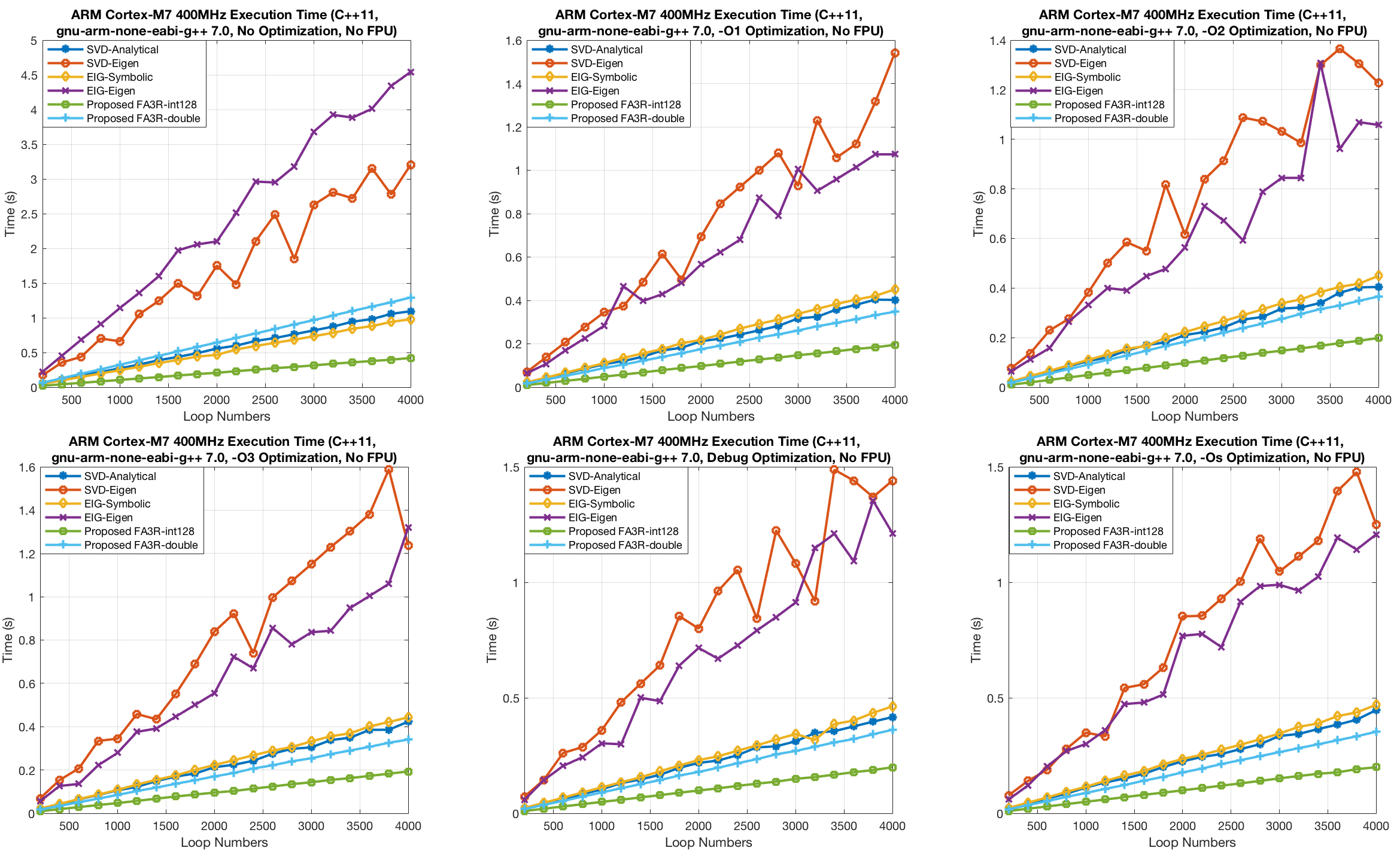}
\caption{Execution time complexity by C++ implementation on the STM32H743VIT6 (Without FPU) containing various compiling optimization parameters and increasing iteration numbers.}
\label{fig:time_ARM_No_FPU}
\end{figure*}

\subsection{Applications}
\setcounter{figure}{9}
\begin{figure}[H]
\centering
\includegraphics[width=0.5\textwidth]{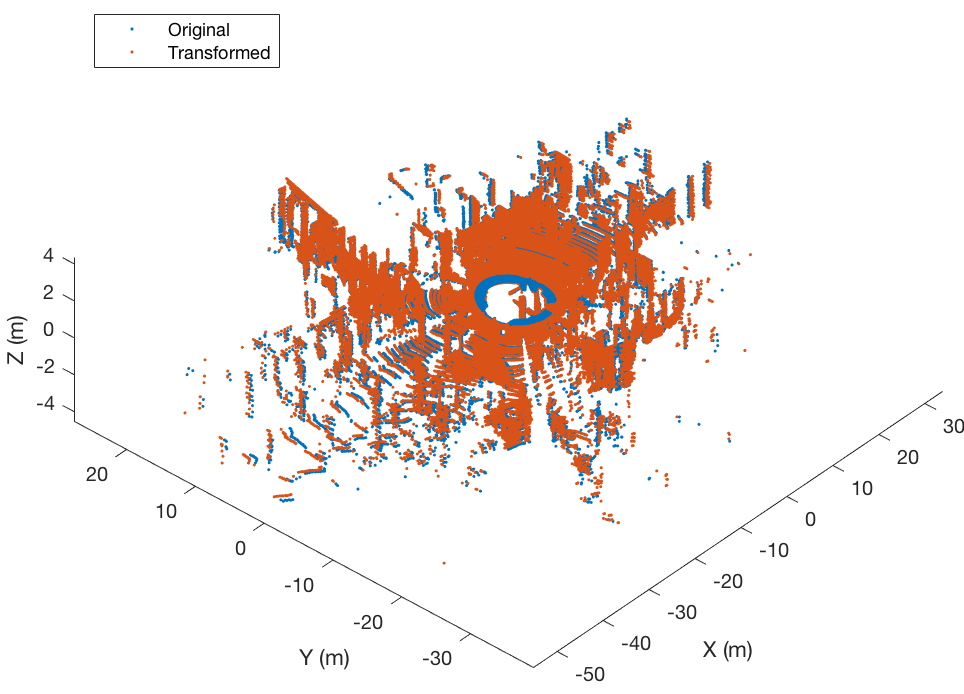}
\caption{The registration performance in scene 1.}
\label{fig:KITTI_FA3R_3D_1}
\end{figure}
In this sub-section, the KITTI dataset \cite{Geiger2013IJRR} is applied to the FA3R for experimental validation. The KITTI dataset contains car-mounted inertial measurement unit (IMU), cameras and a velodyne 3D laser scanner. We use the synced dataset with serial number of '2011\_09\_29\_drive\_0071\_sync' to illustrate the performance.

\setcounter{figure}{10}
\begin{figure}[H]
\centering
\includegraphics[width=0.5\textwidth]{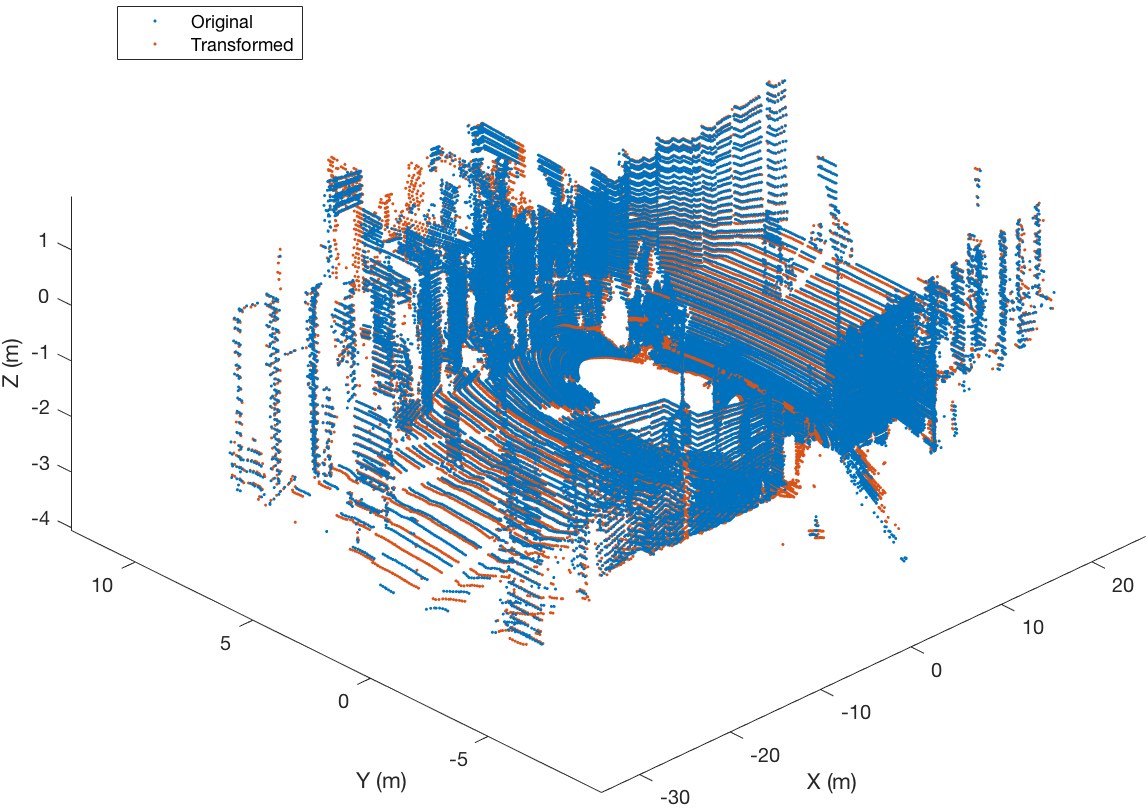}
\caption{The registration performance in scene 2.}
\label{fig:KITTI_FA3R_3D_3}
\end{figure}

The ICP algorithm implemented by MATLAB i.e. the 'pcregrigid' function is utilized for point-cloud registration in which the rigid transform is estimated by the FA3R. Three scenes in the dataset is picked up to show the transformation results. The original point cloud stands for the one that is captured most recently. The transformed one is constructed by the estimated inverse transformation parameters from ICP based on the point cloud in last time instant.

\setcounter{figure}{12}
\begin{figure}[H]
\centering
\includegraphics[width=0.5\textwidth]{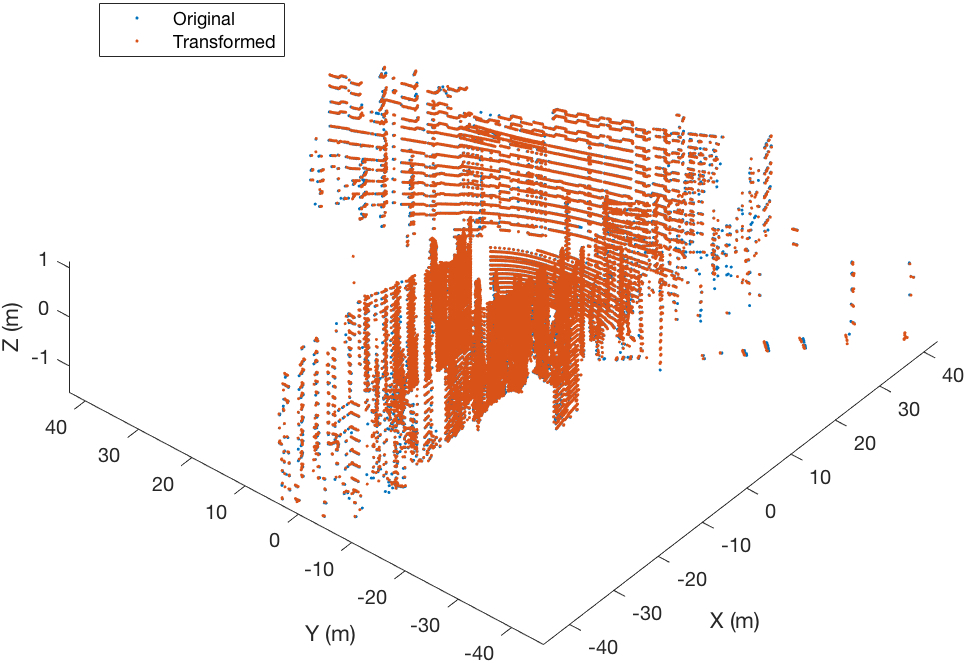}
\caption{The registration performance in scene 3.}
\label{fig:KITTI_FA3R_3D_4}
\end{figure}

The presented results in Fig. \ref{fig:KITTI_FA3R_3D_1}, \ref{fig:KITTI_FA3R_3D_3}, \ref{fig:KITTI_FA3R_3D_4} show that the transformation matrices have been successfully determined using the FA3R without the loss of accuracy compared with SVD. The reconstructed rotation matrix is successively multiplied forming the series of attitude estimates. Along with the ground truth value from the high-precision IMU, the attitude increment results are presented in Fig. \ref{fig:KITTI_attitude}. The SVD and FA3R obtain the same results in real-world scenarios. This verifies that based on the same least-square framework, the FA3R is effective but is easier to be implemented and much more likely to run faster. \\

\setcounter{figure}{13}
\begin{figure*}[ht]
\centering
\includegraphics[width=1.0\textwidth]{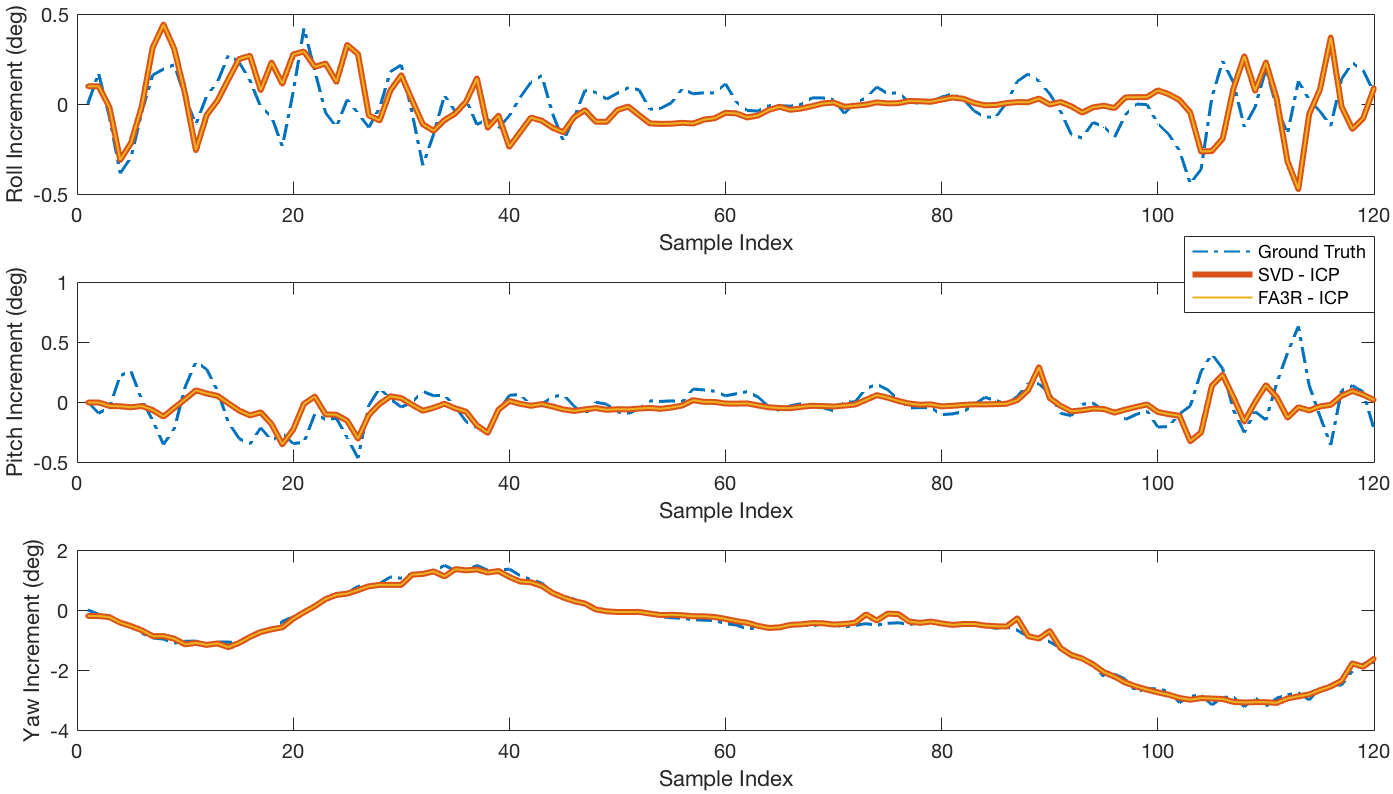}
\caption{The reconstructed incremental attitude results from point cloud registration using SVD and proposed FA3R.}
\label{fig:KITTI_attitude}
\end{figure*}

\setcounter{figure}{15}
\begin{figure*}[hb]
\centering
\includegraphics[width=1.0\textwidth]{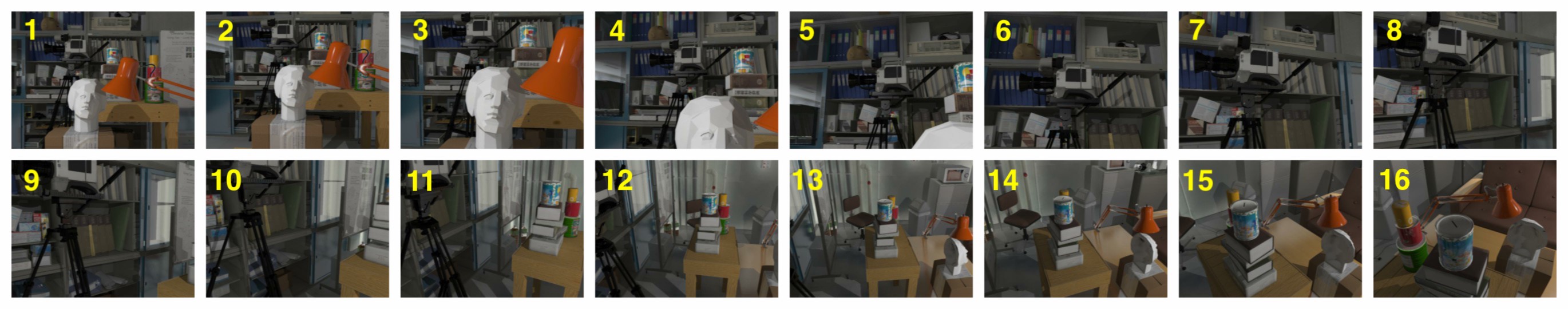}
\caption{The brief description of the 'new tsukuba' video dataset.}
\label{fig:scenes}
\end{figure*}

\indent The proposed FA3R is also very helpful for other problems requiring rigid transform estimation e.g. the Perspective-n-point problem (PnP, \cite{Lepetit2009}). Taking the P3P as example, the FA3R can replace the original SVD for faster computation speed. This allows for the computation of more efficient visual odometry. Here, we select the 'new tsukuba' dataset \cite{Peris2012} from CVLab, Tsukuba University, Japan for performance evaluation. This dataset contains very authentic simulated scenes with ground truth values, which has been extensively used for validation in recent literatures \cite{Zhang2017,Schmeing2015,Ma2017,Clement2018,Alismail2017}. We pick up one small section to compute the trajectories with SVD and FA3R respectively. A brief description in series is shown in Fig. \ref{fig:scenes}. In the calculation, the features are computed with the speeded up robust features (SURF) algorithm. The P3P problem is solved by the analytical algorithm in \cite{Gao2003}. The trajectories are plotted in Fig. \ref{fig:cam_trj}.\\
\indent The results are all the same which verifies the accuracy of the FA3R. But in terms of the formulations of prosed FA3R, the geometry meaning is quite intuitive and the final error exactly denotes the error of rotation. Moreover, the computation time is much lower which will boost its feasibility in engineering applications.
\setcounter{figure}{14}
\begin{figure}[H]
\centering
\includegraphics[width=0.5\textwidth]{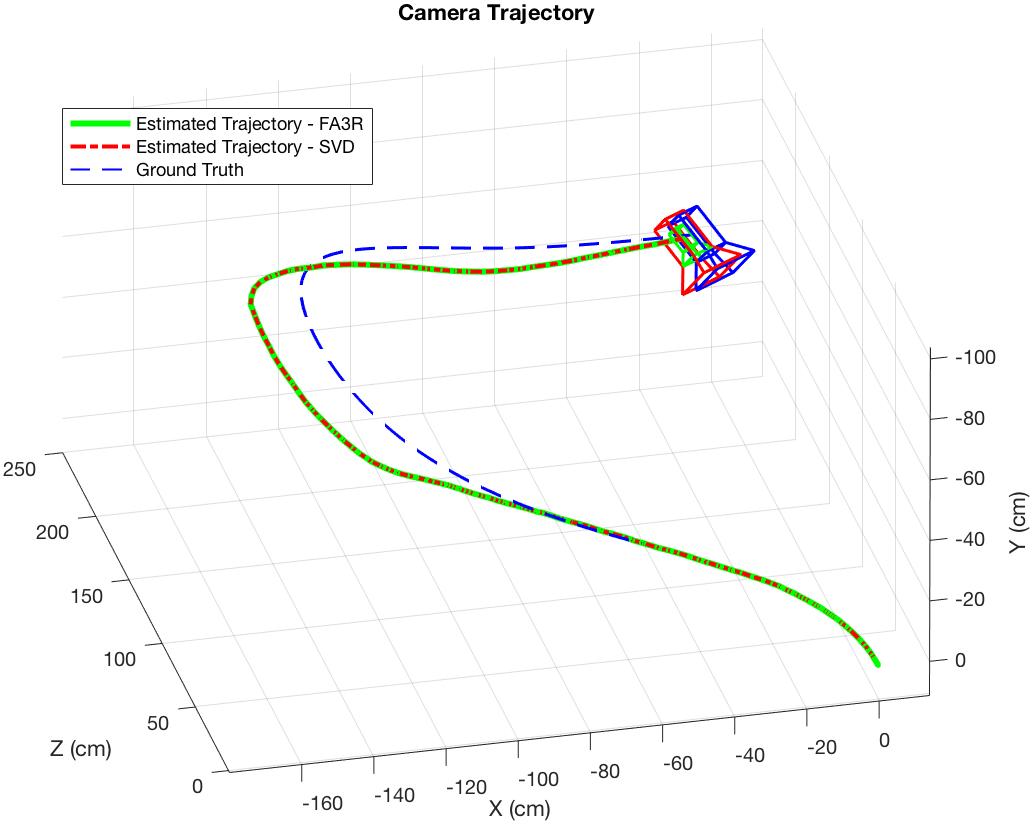}
\caption{The visual odometry's trajectory using SVD and proposed FA3R.}
\label{fig:cam_trj}
\end{figure}

\section{Conclusion}
In this paper, the classical rigid 3D registration problem is revisited. A novel linear simple analytical algorithm is proposed to determine the attitude quaternion. However, different from existing algorithms, the proposed one does not require computation of SVD or eigenvalues. The optimal quaternion can be obtained within very few iteration numbers. The experiments have been conducted to verify the effectiveness of the proposed algorithm. The results show that the proposed FA3R does not have loss on the accuracy and robustness. Rather, the execution time consumption on the PC is improved for $60\% \sim 80\%$. This also gives the practitioners an extremely simple framework to implement on sophisticated platforms like FPGA and GPU. As 3D registration is a very important technique in industrial applications, we hope that this algorithm would benefit related productions in the future. Further works should be dedicated to speed up solving the problem (\ref{iteration}) for faster convergence.

%\begin{figure*}[hb]
%\centering
%\includegraphics[width=1.0\textwidth]{time_ARM_Soft_FPU.jpg}
%\caption{}
%\label{fig:time_ARM_Soft_FPU}
%\end{figure*}

% or


%\appendices
%% you can choose not to have a title for an appendix
%% if you want by leaving the argument blank
%\section{}

\ifCLASSOPTIONcompsoc
  \section*{Acknowledgments}
\else
  \section*{Acknowledgment}
\fi

This research was supported by General Research Fund of Research Grants Council Hong Kong,11210017, in part by Early Career Scheme Project of Research Grants Council Hong Kong, 21202816 and in part by National Natural Science Foundation of China with the grant of No. 41604025. We would like to thank Dr. F. Landis Markley from NASA Goddard Space Flight Center, Dr. Yaguang Yang from U.S. Nuclear Regulatory Commission and Prof. Jinling Wang from the University of New South Wales, for their constructive comments to the contents of this paper.

\ifCLASSOPTIONcaptionsoff
  \newpage
\fi

\end{document}